\documentclass[12pt,preprint]{aastex}

\usepackage{epsfig,graphicx,amssymb,epstopdf,mathrsfs}
\newcommand{\Swift}{{\em Swift}}
\newcommand{\swift}{\textit{Swift}}

\newcommand{\bmm}[1]{\mathbf{#1}}

\def\deg{\hbox{$^\circ$}}

\def\amin{\hbox{$^\prime$}}
\def\arcsec{\hbox{$^{\prime\prime}$}}
\def\farcm{\hbox{$.\!^{\prime}\hspace{-0.08em}$}}  % Fractions of arcminutes
\def\farcs{\hbox{$.\hspace{-0.25em}^{\prime\prime}\hspace{-0.08em}$}}  % Fractions of arcseconds
\def\fss{\hbox{$.\!\!^{\rm s}$}}  %fractions of seconds

\def\uJy{\hbox{$\mu$Jy}}
\def\simlt{\mathrel{\hbox{\rlap{\hbox{\lower4pt\hbox{$\sim$}}}\hbox{$<$}}}}
\def\simgt{\mathrel{\hbox{\rlap{\hbox{\lower4pt\hbox{$\sim$}}}\hbox{$>$}}}}

\def\h{$^{\rm h}$}
\def\m{$^{\rm m}$}

\newcommand{\Msun}{\mbox{$M_\odot$}}
\newcommand{\Zsun}{\mbox{$Z_\odot$}}

\newcommand{\etal}{\mbox{et al.}}

\newcommand{\newtext}[1]{\textbf{#1}}

\newcommand{\nbytx}[2]{#1$\times$\,#2}

\newcommand{\usnob}{\mbox{USNO-B1.0}}

\newcommand{\peryear}{{{\rm yr}$^{-1}$}}

\newcommand{\rp}{\mbox{$r^\prime$}}
\newcommand{\ip}{\mbox{$i^\prime$}}
\newcommand{\zp}{\mbox{$z^\prime$}}

\newcommand{\Mrfuv}{\mbox{$M_{\rm UV}$}}
\newcommand{\zmean}{\mbox{$\langle z \rangle$}}

\newcommand{\Msupp}{\mbox{$M_{\rm supp}$}}

\newcommand{\Msunpyr}{\Msun\,\peryear}

\def\fm{\hbox{$.\!\!^{\rm m}$}}            % Fractions of magnitudes

\begin{document}

\title{Gamma-Ray Bursts Trace UV Metrics of Star Formation over $3 < z <
  5$\thanks{Partly based on observations collected at the European
    Organisation for Astronomical Research in the Southern Hemisphere
    under IDs 089.A-0120(A) and 091.A-0786(A).}}

\author{J.~Greiner\altaffilmark{1,2},
        D.~B.~Fox\altaffilmark{3},
        P.~Schady\altaffilmark{1},  
        T.~Kr\"uhler\altaffilmark{4},
        M.~Trenti\altaffilmark{5,6},
        A.~Cikota\altaffilmark{7},
        J.~Bolmer\altaffilmark{1,8},
        J.~Elliott\altaffilmark{1,9},
        C.~Delvaux\altaffilmark{1},
        R.~Perna\altaffilmark{10},
% alphabetical from here
        P.~Afonso\altaffilmark{11},
        D.~A.~Kann\altaffilmark{12},
        S.~Klose\altaffilmark{12},
        S.~Savaglio\altaffilmark{13,1},
        S.~Schmidl\altaffilmark{12},
        T.~Schweyer\altaffilmark{1,8},
        M.~Tanga\altaffilmark{1},
        K.~Varela\altaffilmark{1}
} 

\email{jcg@mpe.mpg.de, 
       dfox@psu.edu,
       pschady@mpe.mpg.de,
       tkruehle@eso.org,
       michele.trenti@unimelb.edu.au,
       aleksandar.cikota@student.uibk.ac.at,
       jan@bolmer.de,
       jonathan.elliott@cfa.harvard.edu,
       delvaux@mpe.mpg.de,
       rosalba.perna@stonybrook.edu,
       afonsop@arc.losrios.edu,
       klose@tls-tautenburg.de,
       sandra.savaglio@googlemail.com,
       schmidl@tls-tautenburg.de,
       mpe@welterde.de,
       mohit@mpe.mpg.de,
       kvarela@mpe.mpg.de \\}

\altaffiltext{1}{Max Planck Institute for Extraterrestrial Physics,
  Giessenbachstrasse, 85748 Garching, Germany}

\altaffiltext{2}{Excellence Cluster Universe, Technische Universit\"{a}t 
      M\"{u}nchen,  Boltzmannstra{\ss}e 2, 85748, Garching, Germany}

\altaffiltext{3}{Department of Astronomy \& Astrophysics, Center for
  Theoretical and Observational Cosmology, and Center for Particle \&
  Gravitational Astrophysics, Pennsylvania State University,
  University Park, PA 16802, USA}

\altaffiltext{4}{European Southern Observatory, Alonso de C\'{o}rdova 3107, 
            Vitacura, Casilla 19001, Santiago 19, Chile}

\altaffiltext{5}{Institute of Astronomy and Kavli Institute for Cosmology, 
   University of Cambridge, Madingley Road, Cambridge, CB3 0HA, United Kingdom}

\altaffiltext{6}{School of Physics, University of Melbourne,
      VIC 3010 Australia}

\altaffiltext{7}{Institut f\"ur Astro- und Teilchenphysik, Universit\"at 
      Innsbruck, Technikerstrasse 25/8,  6020 Innsbruck, Austria}

\altaffiltext{8}{Technische Universit\"at M\"unchen, Physik Dept.,
              James-Franck-Str., 85748 Garching, Germany}

\altaffiltext{9}{Astrophysics Data System, Harvard-Smithsonian Center
  for Astrophysics, 60 Garden St., Cambridge, MA 02138, U.S.A.}

\altaffiltext{10}{Dept. of Physics and Astronomy, Stony Brook University,
          NY 11794-3800, USA}

\altaffiltext{11}{American River College, Physics \& Astronomy Dept., 
          4700 College Oak Drive, Sacramento, CA 95841, USA}

\altaffiltext{12}{Th\"uringer Landessternwarte Tautenburg, Sternwarte 5,
             07778 Tautenburg,  Germany}

\altaffiltext{13}{Universita della Calabria, 87036 Arcavacata di Rende,
          via P. Bucci, Italy}

\shorttitle{GRBs Trace Star Formation over $3<z<5$}
\shortauthors{Greiner \etal }

%%%%%%%%%%%%%%%%%%%%%%%%%%%%%%%%%%%%%%%%

\begin{abstract}
We present the first uniform treatment of long duration gamma-ray
burst (GRB) host galaxy detections and upper limits over the redshift
range $3<z<5$, a key epoch for observational and theoretical efforts
to understand the processes, environments, and consequences of early
cosmic star formation. We contribute deep imaging observations of 13
GRB positions yielding the discovery of eight new host galaxies. We
use this dataset in tandem with previously published observations of
31 further GRB positions to estimate or constrain the host galaxy
rest-frame ultraviolet (UV; $\lambda=1600$\AA) absolute magnitudes
\Mrfuv. We then use the combined set of 44 \Mrfuv\ estimates and
limits to construct the \Mrfuv\ luminosity function (LF) for GRB host
galaxies over $3<z<5$ and compare it to expectations from Lyman break
galaxy (LBG) photometric surveys with the Hubble Space
Telescope. Adopting standard prescriptions for the luminosity
dependence of galaxy dust obscuration (and hence, total star formation
rate), we find that our LF is \newtext{compatible} with LBG
observations over a factor of 600$\times$ in host luminosity, from
$\Mrfuv = -22.5$\,mag to $> -15.6$\,mag, and with extrapolations of
the assumed Schechter-type LF well beyond this range. We review
proposed astrophysical and observational biases for our sample, and
find they are for the most part minimal.  We therefore conclude, as
the simplest interpretation of our results, that GRBs successfully
trace UV metrics of cosmic star formation over the range $3<z<5$. Our
findings suggest GRBs are providing an accurate picture of star
formation processes from $z\approx 3$ out to the highest redshifts.
\end{abstract}

% Up to 6 keywords

\maketitle

%%%%%%%%%%%%%%%%%%%%%%%%%%%%%%%%%%%%%%%%

\section{Introduction}
\label{sec:intro}

A fundamental goal of modern cosmology is to understand the history of
star formation in our Universe, from its earliest epochs to the
present day. Efforts over recent decades, including ambitious
ground-based spectroscopic surveys and space-based observations in the
ultraviolet (GALEX), optical to near-infrared (Hubble Space
Telescope), and longer wavelengths (Spitzer, Herschel) have enabled
multiple determinations and cross-checks, including extinction
corrections, of the cosmic star formation rate (SFR) at redshifts
$z\simlt 3$, and the first credible estimates over $3\simlt z \simlt
8$; for a recent review and references, see \citet{md14}.

Star formation in the early Universe, at $z>3$, is of particular
interest as it tracks the formation of the first galaxies, which
should account for the redshift and timescale of the $z\approx 6$
cosmic reionization and may yield insight into the dark matter-driven
formation of the first cosmic structures
\citep[e.g.,][]{ysh+03,tsb+10,th10,dw12,jcn+12,tpt13}. Accurately
quantifying star formation at these redshifts presents obvious
challenges: at these distances even the most luminous galaxies are
faint and nearly inaccessible to spectroscopic study; moreover,
high-sensitivity observations are currently (until the advent of the
James Webb Space Telescope, JWST) restricted to rest-frame ultraviolet
(UV) and optical bands. Nonetheless, the underlying scientific promise
has motivated intense observational efforts.

%%%%%%%%%%%%%%%%%%%%%%%%%%%%%%%%%%%%%%%%

% Observational efforts

Multiple deep surveys with the Hubble Space Telescope (HST) have
collected over 10,000 photometrically-selected Lyman break galaxies
(LBGs) over $3 \simlt z \simlt 8$
\citep{bil+11,obi+12,emd+13,sre+13,Schmidt+14,bio+14,bio+15}, with
multiband imaging over substantial regions allowing the fitting of full
spectral energy distributions (SEDs), and thus stellar population
models, for individual galaxies \citep[e.g.,][]{dcm+14}. These surveys
have been complemented by HST observations of the most massive known
galaxy clusters, which take advantage of the gravitational-lensing
boost to extend $z>6$ LBG luminosity functions (LFs) by a full
magnitude \citep{ark+14}, and have yielded LBG candidates out to
$z\approx 9.6$ \citep{zpz+12,bbz+14}.

Meanwhile via ground-based facilities, narrow-band surveys targeting
emission-line galaxies have collected substantial samples over $5<z<7$
\citep{skd+06,osf+10,oik+10}; cosmic microwave background (CMB)
experiments have discovered dozens of gravitationally-lensed, dusty,
star-forming galaxies at $z>4$ \citep{vcs+10,mcv+13,mgm+14}; and
large-area near-infrared (NIR) surveys are yielding competitive
constraints on the bright end of the $z\approx 7$ galaxy LF
\citep{bdm+14} and pushing the frontier of quasar discovery out to
$z>7$ \citep{mwv+11}.

% Current status

With the latest results from the Planck Collaboration
\citeyearpar{planck15_13} yielding a reduced estimate for the Thomson
scattering optical depth to the CMB ($\tau_{\rm CMB}=0.066\pm 0.012$)
consistent with reionization ending at $z\approx 6$
\citep{sht+12,ref+15,bio+15b}, the chief remaining uncertainty in our
estimates of high-redshift SFR must be the quantity of star formation
happening in faint galaxies, beyond the reach of even the deepest HST
fields. Such galaxies are too faint for current galaxy surveys; at
higher redshifts, many have flux densities $f_\nu \simlt 1$\,nJy that
will challenge even JWST, and thus most conceivable observational
tests.

%%%%%%%%%%%%%%%%%%%%%%%%%%%%%%%%%%%%%%%%

% GRBs for star formation

In this respect, long-duration gamma-ray bursts (GRBs) offer an
elegant and complementary approach, currently reaching to $z\simgt 8$
\citep{tfl+09,sdc+09,clf+11}, that could be extended to $z\simgt 12$
with relatively modest space-based observatories \citep{brf+10,
  pwb+11, gma+12}.  Due to their link to core-collapse supernovae
(SNe) \citep{smg+03,hsm+03}, GRBs probe the formation of massive
stars.  Intrinsically luminous and dust-penetrating, GRBs accurately
pinpoint regions of active SF -- in three dimensions -- independent of
galaxy luminosity, dust obscuration, and precise redshift.

% Status of GRBs for SFR

Past efforts to use GRBs as tracers of cosmic star formation (SF) have
compared the GRB redshift distribution to other independent
estimators. Since SF at low redshifts is considered
well-characterized, the aim of these studies has been to validate or
calibrate the GRB distribution over $z\simlt 3$ so as to leverage
knowledge of the GRB redshift distribution at higher redshifts,
$3\simlt z\simlt 8$, to infer cosmic SFR and explore the consequences
for cosmic reionization. Initial efforts using the
\swift\ \citep{swift} GRB redshift distribution found that the GRB
rate rose more rapidly with redshift over $z\simlt 4$ than other
estimators \citep{ykb+08,kyb+09}. Adopting the implied correction,
$(1+z)^\alpha$ with $0.6 < \alpha < 1.8$, led to $z>4$ SFR predictions
consistent with extrapolations of the LBG LF at those redshifts
\citep{kyb+09}.

\citet{wp10} also found a steep rise in the GRB rate at $z\simlt 3$
and a relatively slow decay at $z>4$, which they suggested was
inconsistent with LBG SFR estimates. However, their derived $z>4$ SFR
(Fig.~9) closely resembles both the evolution-corrected GRB-based SFR
and the ``LF integrated'' LBG-based SFR from \citet[][their
  Fig.~4]{kyb+09}. Given the uncertainties, it may be accurate to
characterize these initial papers as demonstrating that the $z>4$ SFRs
implied by GRB rates and LBG surveys are consistent over a range of
plausible LBG LF extrapolations and GRB evolution corrections.

This \newtext{consistency of GRB- and LBG-derived SFRs would be 
  in accord}
with \citet{re12}, who found that varying their treatment of the
unknown redshifts of dark bursts (GRBs that lack bright optical
afterglows, and are hence missing from redshift samples based on
afterglow spectroscopy) had a significant impact on the strength of
the evolution correction needed to bring GRB and other SFR metrics
into agreement at $z<4$: putting all dark bursts at their maximum
likely redshifts suggested a mild GRB evolution correction,
$\alpha\approx 0.5$; while putting all dark bursts at low redshifts
favored no correction, $\alpha\approx 0$.  Along these lines, we note
that luminosity-based selection of GRBs, pursued in some cases to
improve sample completeness, has been shown to affect the strength of
the inferred anti-metallicity bias \citep{egk+12}.

If differential evolution of the GRB rate and SFR is required at
low-redshift, there is a broad consensus that this would reflect an
anti-metallicity bias in the GRB population, thanks to past
suggestions on both theoretical \citep[e.g.,][]{wh06,yln06} and
observational \citep[e.g.,][]{sgb+06,fls+06} grounds. However, since
the average metallicity of star-forming galaxies drops precipitously
from $z=0$ to $z\approx 3$ \citep{mng+08}, even if anti-metallicity
bias is present the implied correction from GRB rates to SFR at
$z\simgt 3$ might be modest \citep{tpt13}.  Indeed, comparison of
high-resolution cosmology simulations to the observed GRB sample 
suggests $\alpha\approx 0$ at redshifts of $z>5$ \citep{ekg+15}.

%%%%%%%%%%%%%%%%%%%%%%%%%%%%%%%%%%%%%%%%%%%%%%%%%%

% Paper outline

In this paper, we seek to make a different type of global comparison
between GRBs and UV-based star formation metrics, calculating the UV
luminosity function of GRB host galaxies over $3<z<5$ and comparing it
to prior expectations from LBG surveys 
\newtext{(see also \citealt{sch+15})}. 
We focus our efforts on this redshift range because:
(1) The accuracy and completeness of current SFR estimates at these
redshifts, in terms of extinction corrections and the necessary
extrapolation to unobserved faint galaxies, remain subject to active
debate;
(2) At these redshifts any metallicity bias in the selection of GRB
host galaxies is likely to be reduced or, potentially, negligible;
(3) A sufficient number of GRB redshifts have been measured within
this range to make statistical analyses useful; and
(4) Large-aperture ground-based telescopes (and HST) can readily
detect host galaxies of modest luminosity at these redshifts or,
alternatively, yield useful constraints on their magnitudes.
Naturally, we carry out this work in hopes that it will shed light on
the connection between GRBs and SF not just within, but also beyond,
our targeted redshift range.

% Table 1:  13 host galaxies, 15 observations
%              9 detections of 8 galaxies (1 galaxy detected twice)
%              6 limits (5 galaxies undetected; 1 galaxy with two limits)

% Added:    31 host galaxies, 61 observations
%             35 detections of 21 galaxies
%             26 limits (10 galaxies undetected)

In Section~\ref{sec:obs} below we present 15 observations of 13 GRB
host galaxies over $3<z<5$, which yield detections of eight and deep
limits of five GRB hosts.  We supplement these data with 61
observations of 31 additional host galaxies from the literature,
yielding detections of 21 and limits of 10 hosts.  With a uniform
treatment of the new and previously published data, informed by the
results of the HST LBG surveys, we convert each observation into a
rest-frame UV ($\lambda = 1600$\AA) absolute magnitude \Mrfuv\ or
limit, using a $\Lambda$CDM cosmology with $H_{\rm o}$=70 km/s/Mpc,
$\Omega_{\rm M}$=0.3, $\Omega_{\rm \Lambda}$=0.7.
In Section~\ref{sec:interpret} we construct the UV LF for the GRB host
galaxies, and compare it to the SFR-weighted LF of Lyman-break
galaxies in this redshift range. We then discuss the possible
influence of proposed astrophysical biases, including an anti-metallicity
bias in GRB production and suppression of star-formation in low-mass
halos, and observational selection effects, including interloper
galaxies, ``dark bursts'' potentially missing from our sample, and
publication bias.
Our conclusions are presented in Section~\ref{sec:conclude}.

%%%%%%%%%%%%%%%%%%%%%%%%%%%%%%%%%%%%%%%%
%%%%%%%%%%%%%%%%%%%%%%%%%%%%%%%%%%%%%%%%

\section{Observations}
\label{sec:obs}

Presently, measured redshifts for GRB afterglows are known for more
than 400 GRBs\footnote{see the online collection at
  \url{http://mpe.mpg.de/$\sim$jcg/grbgen.html}}, among those 65 (as
of February 2015) in the redshift range $3<z<5$. Of those, host
measurements for 31 are reported in the literature and in the GRB Host
Studies (GHostS) database\footnote{GHostS:
  \url{http://www.grbhosts.org/}}, at widely disparate sensitivities.
We add our own data, in particular FORS2/VLT \citep{ar92} observations
of 7 GRB hosts, and GROND/2.2m \citep{gbc+07, gbc+08} observations for
another 6 GRB hosts.

Our VLT observations of GRB host galaxies at $3 < z < 5$ are
summarized in Table~\ref{tab:obs}. FORS2/$R$-band observations are taken as
a sequence of 1140\,s integrations, while FORS2/$I$-band observations are
taken as a sequence of 240\,s integrations, owing to the brighter sky
in that bandpass. Observations were carried out in service mode,
extending over two or more nights in each case, and were almost all of
high quality. The raw data are available for downloading from the ESO
VLT data archive\footnote{VLT data archive:
  \url{http://dataportal.eso.org}}. Pointings were chosen to put the
targeted region of sky in a clean portion of CCD~2 on FORS2, southwest
of the pointing center. 

GROND observations (also listed in Table~\ref{tab:obs}) were taken as
a sequence of 369\,s exposures using four or six telescope dither
positions.  Imaging was done in all seven GROND filters
simultaneously; for each target we report the observed magnitude for 
the bluest filter redward of, and not affected by, Ly-$\alpha$.

%%%%%%%%%%%%%%%%%%%%%%%%%%%%%%

\subsection{Data Reduction}
\label{sub:reduction}

After bias-subtraction and flat-fielding, VLT images are trimmed to
exclude the vignetted corners of the FORS2 field of view; CCD~1 images
are trimmed to a 3\farcm53 by 2\farcm01 rectangular region, and CCD~2
images to a 3\farcm53 by 1\farcm50 rectangular region.

Individual CCD images are processed with SExtractor \citep{ba96} in
order to identify and mask bright objects and fit and subtract a
smoothly-varying model of the sky background. We observe mild fringing
at a level of $<$1\% of the sky background which is not reliably
repeated from one observing epoch to the next. For this reason, we
construct an object-masked fringe image for each observing epoch and
scale and subtract it from the images before proceeding. Using the
native WCS, the two CCD images are then integrated into a single image
with an 18-pixel (2\farcs28) gap between the two fields.

Coaddition of the multiple frames for each field proceeds via a
two-step process. In the first step, images are aligned via
cross-correlation analysis and a median stack is constructed. This
median image is used to identify bad pixels and cosmic ray-affected
pixels in each individual frame, updating the previous (static) bad
pixel masks for each detector. Masked pixels in each image are
replaced by corresponding values from the median image, and the images
are re-aligned with a second cross-correlation analysis. The final
coadded image is then generated as the sigma-clipped mean of the
masked and aligned individual frames. The coadded image is trimmed to
the size of the best-centered individual frame, yielding a single
3\farcm53 by 3\farcm55 image with roughly constant depth across
the field. We then refine the WCS of this final image by reference to
the SDSS (when available) or \usnob\ catalogs; the resulting WCS solutions
have uniformly acceptable residuals of $<$0\farcs2 (rms).

These residuals are due to the mapping against the catalog positions.
Astrometry between two GROND images (for afterglow and host, respectively)
or between GROND and FORS2 is even more accurate. In order to
determine the position of the afterglow relative to the host galaxy,
the following procedure is used. First, a source detection is run
over the GROND images containing the afterglow, providing a list of 
x,y coordinates of the afterglow and neighbouring field stars (extended
objects were de-selected). This list is then used as ``catalog'' input
list for the astrometry of the FORS2 images. This provides relative
accuracies of $<$0\farcs05 and $<$0\farcs1 (rms) for GROND-internal
and GROND-FORS2 mapping, respectively. In Fig.~\ref{fig:obs} the position
of the afterglow is plotted with a cross, while the circle denotes the
aperture used to extract the host photometry.

Photometric analysis is done using standard IRAF tasks~\citep{iraf93},
as documented in full in \cite{kkg+08}. In short, bright
stars within the field are used to fit a point-spread function (PSF)
which then is applied to the entire field.  Absolute photometric
calibration of the FORS images was achieved with GROND by observing a
Sloan digital Sky Survey (SDSS) field~\citep{sdss8} that was closest
to the host galaxy, and consecutively the corresponding host field (in
few cases this was not needed as the GRB lies already in SDSS-covered
area).

%Simple aperture photometry is used for the host galaxy.
% Host Photometry

With accurate afterglow positions registered on our VLT and GROND
frames, and using the well-known $<$1\arcsec\ offsets of long-duration
GRBs from their hosts \citep{bkd+02}, we are finally in the position
to search for a host related to the corresponding GRB afterglow.
Simple aperture photometry is used for the host galaxy.  For each
galaxy a curve-of-growth analysis was carried out to determine the
aperture size for which \newtext{the signal-to-noise ratio is
  optimized (also added as separate column in Table~\ref{tab:obs})}.
%all of the emission was obtained. 
%The aperture size used for each galaxy is shown in Fig.~\ref{fig:obs}
%{\bf and in Table~\ref{tab:obs}. 
\newtext{We also list the centroid positions of the host galaxies
in Table~\ref{tab:obs}. }  
A selection of stars for each host field were then used to determine the 
associated zero point. 

%%%%%%%%%%%%%%%%%%%%%%%%%%%%%%%%%%%%%%%%%%%%%%%%%%

\subsection{Results and Source Notes}
\label{sub:results}

Fig.~\ref{fig:obs} shows the final image of each of our sample GRBs.
Table~\ref{tab:obs} contains the details of the observations,
including the GRB name (column 1), the redshift (2),
telescope/instrument (3), filter (4), exposure (5), \newtext{centroid
  position of the host (0\farcs3 uncertainty) (6), the aperture size
  as determined from the curve-of-growth analysis (7)} and the host
magnitude with uncertainty, reporting a 2$\sigma$ upper limit for
non-detections (8).  \newtext{The three GRBs with explicit errors in
  the redshift column have photometric redshifts.}

%%%%%%%%%%%%%%%%%%%%%%%%%%%%%%%%%%%%%%%%

\begin{figure}[p]
\vspace*{-0.3cm}
% 080810 090313 090516 090519 091109 100513 100518 110818A 120909A 120922A 121201A 130408A
% 3 in a row: width=5.4 | 2 in row: width=8.1cm
\includegraphics[width=7.4cm]{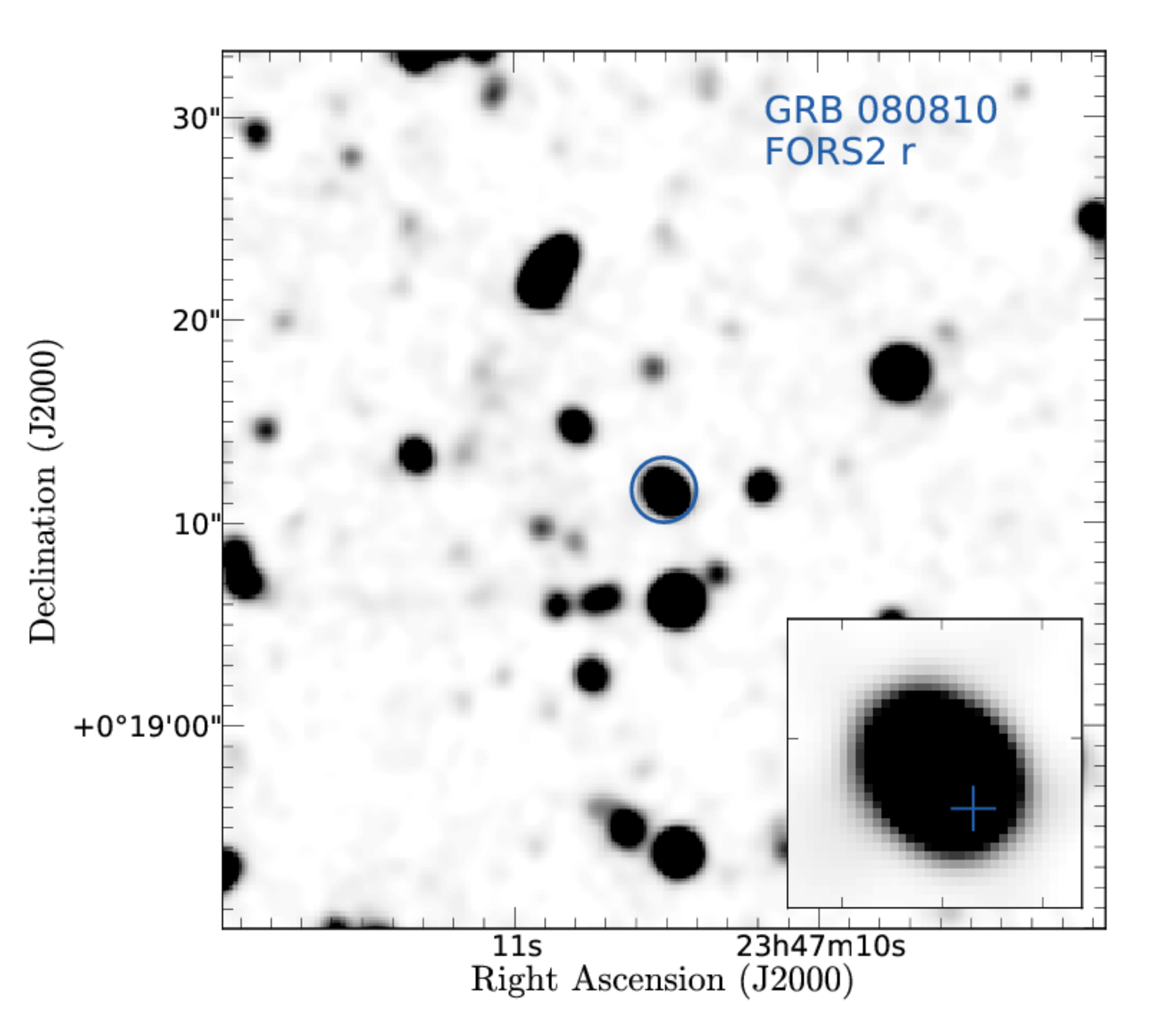}
\includegraphics[width=7.4cm]{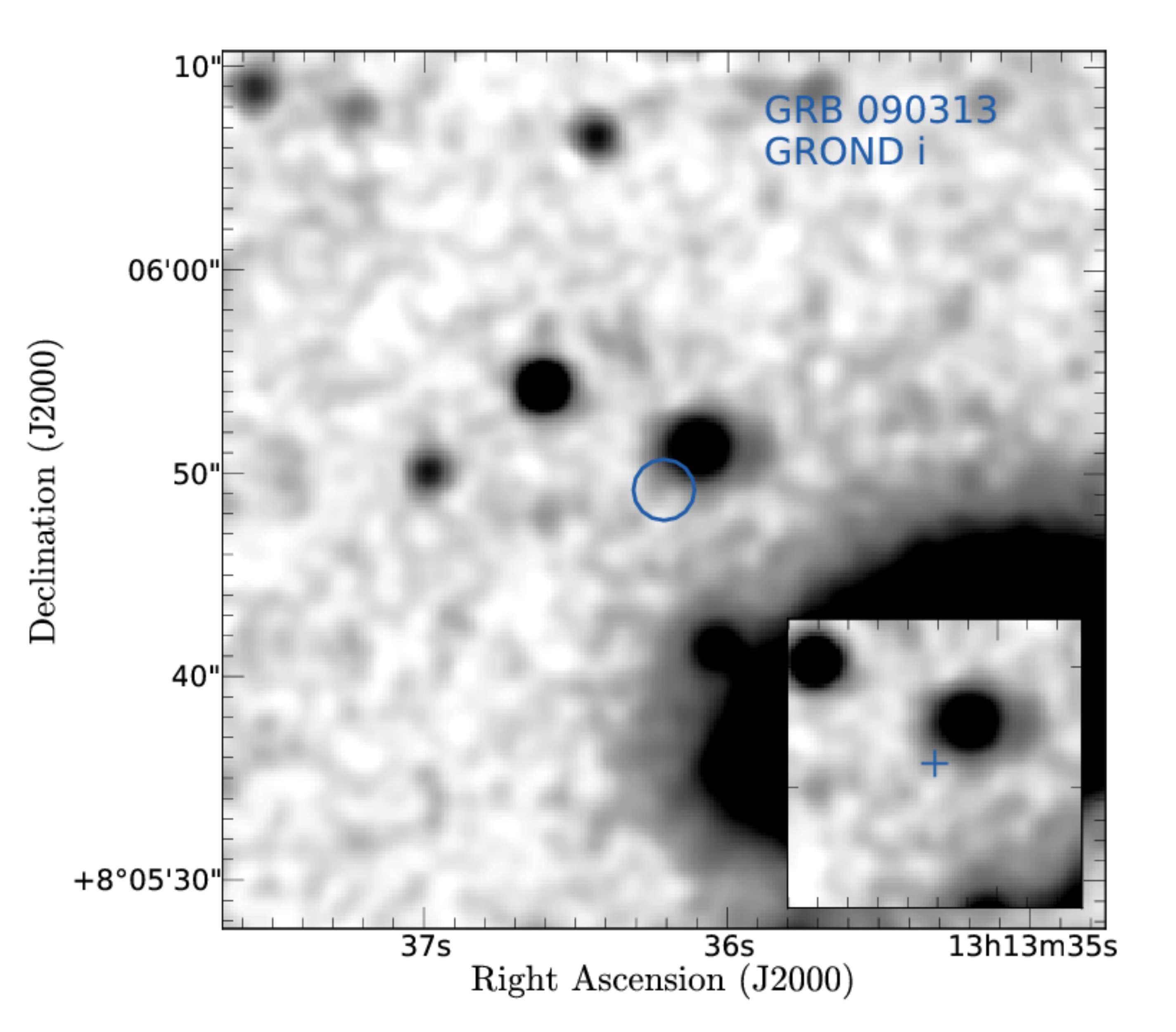} \\ 
\includegraphics[width=7.4cm]{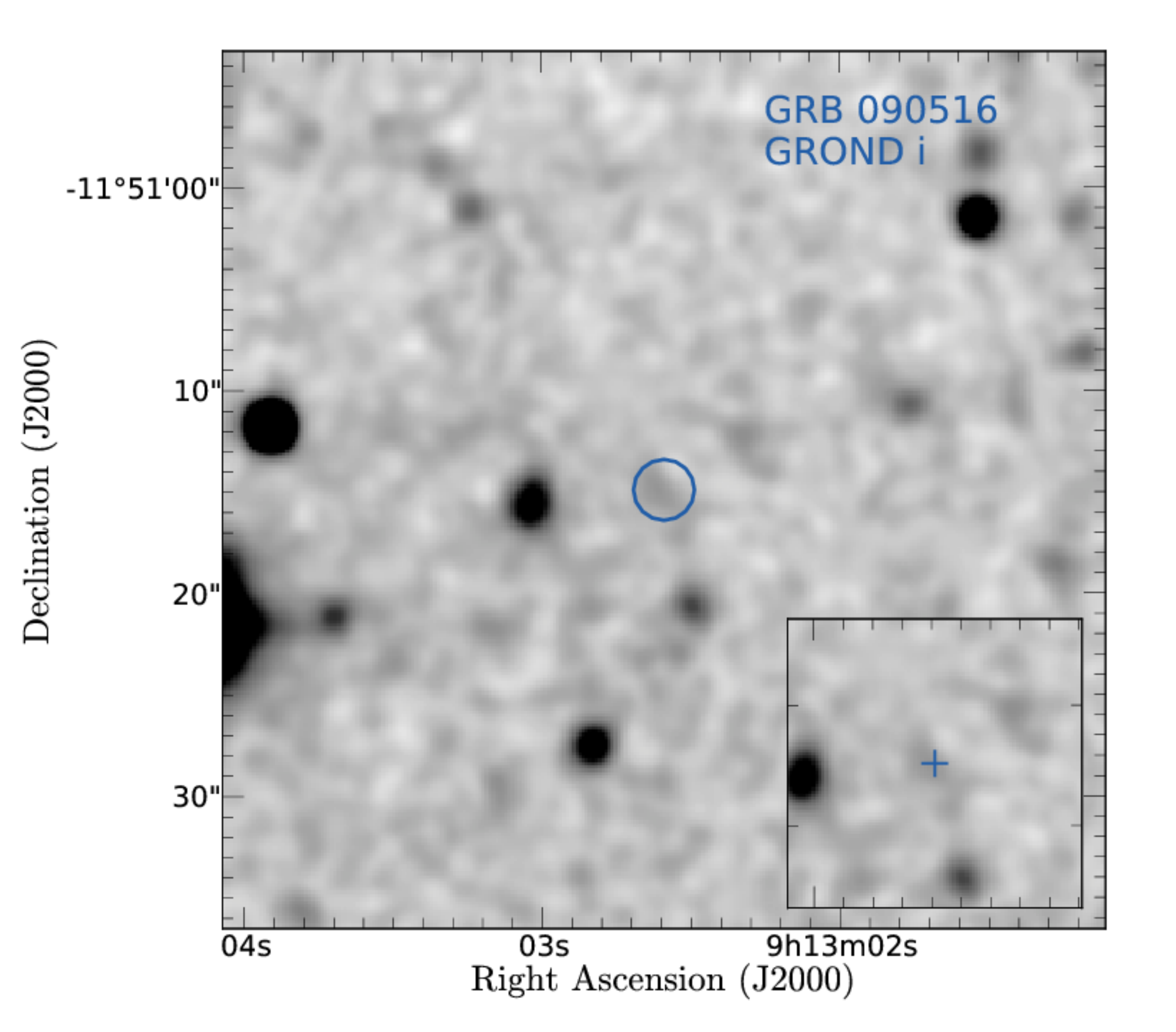} 
\includegraphics[width=7.4cm]{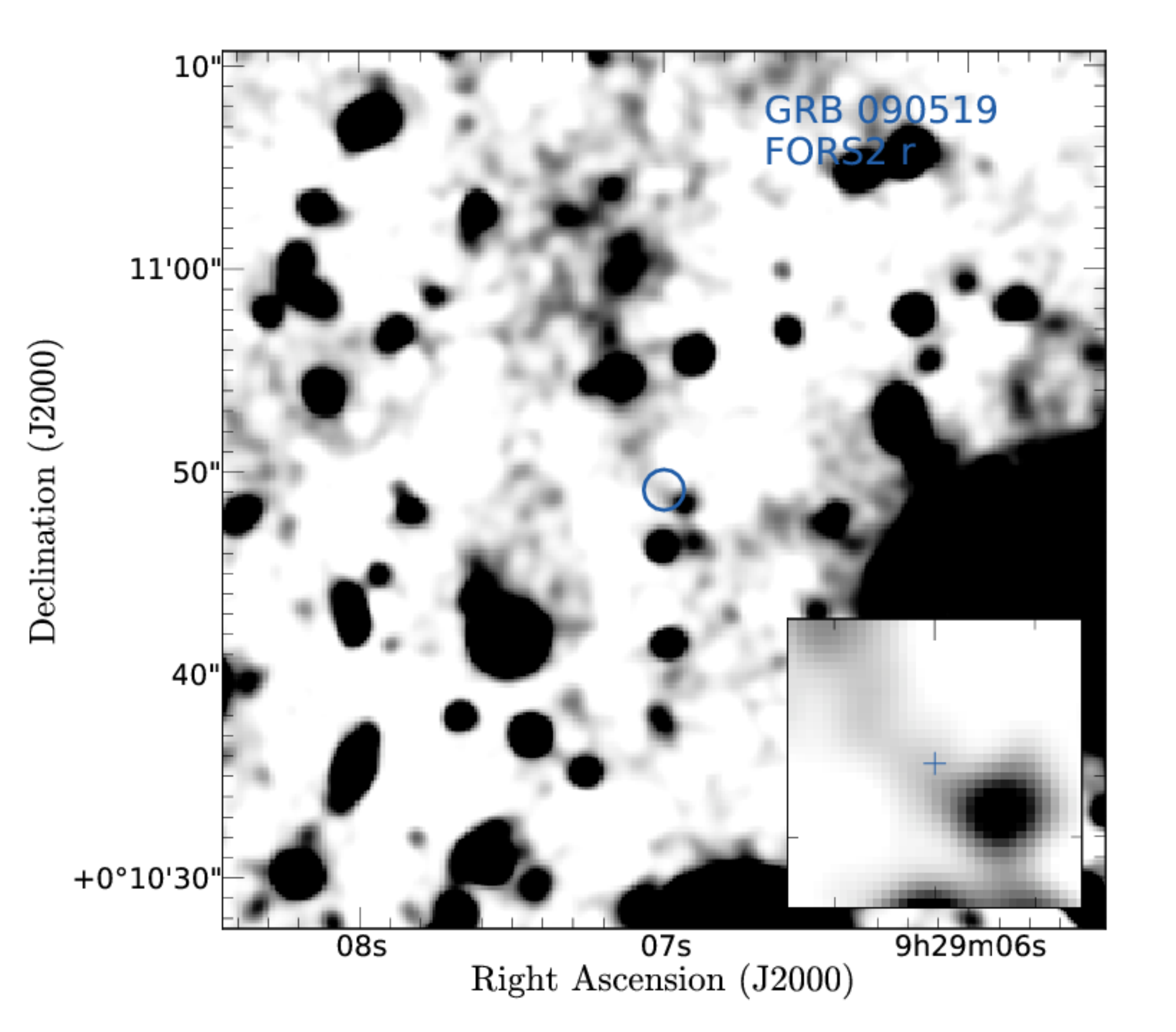} \\
\includegraphics[width=7.4cm]{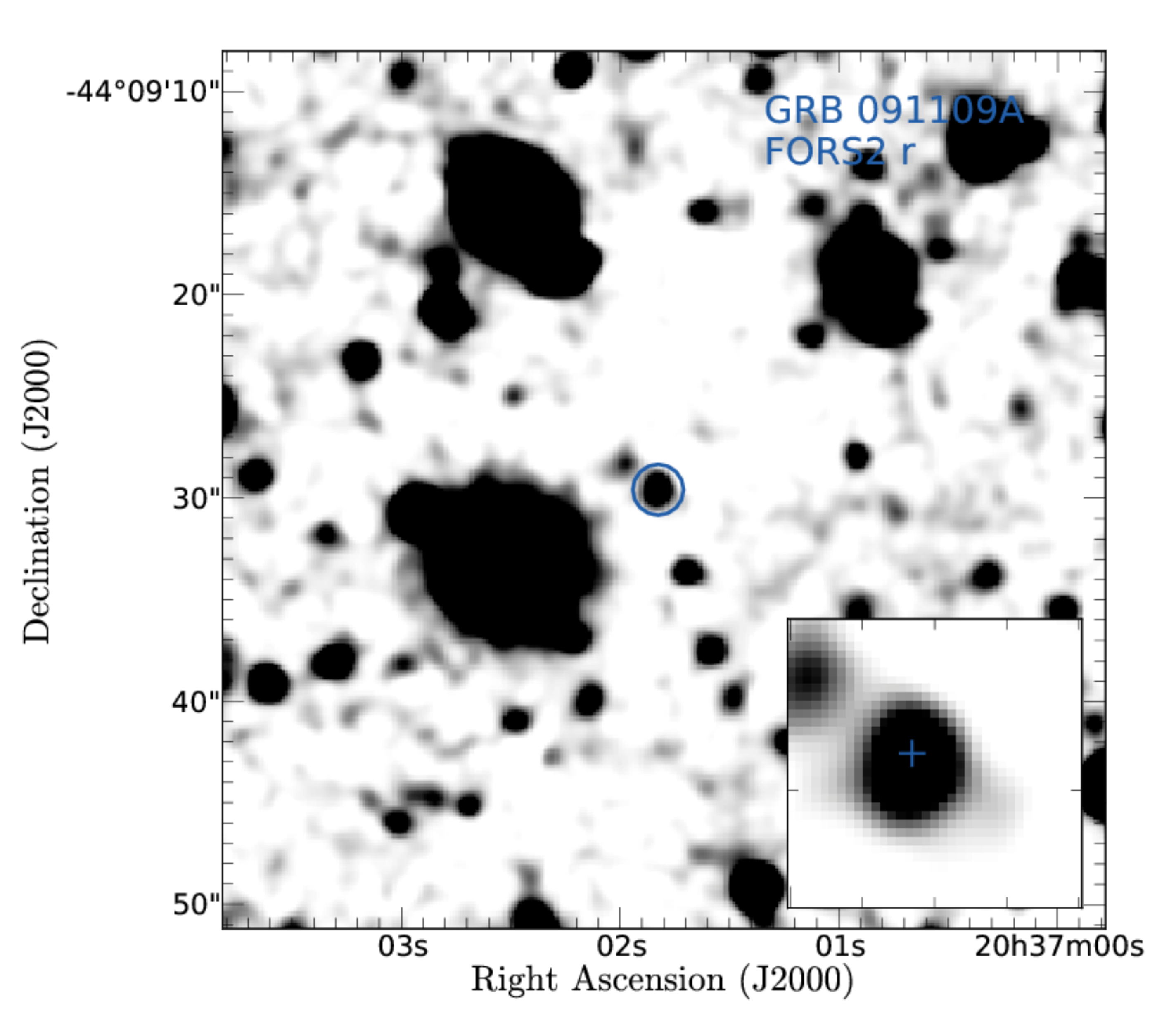}
\includegraphics[width=7.4cm]{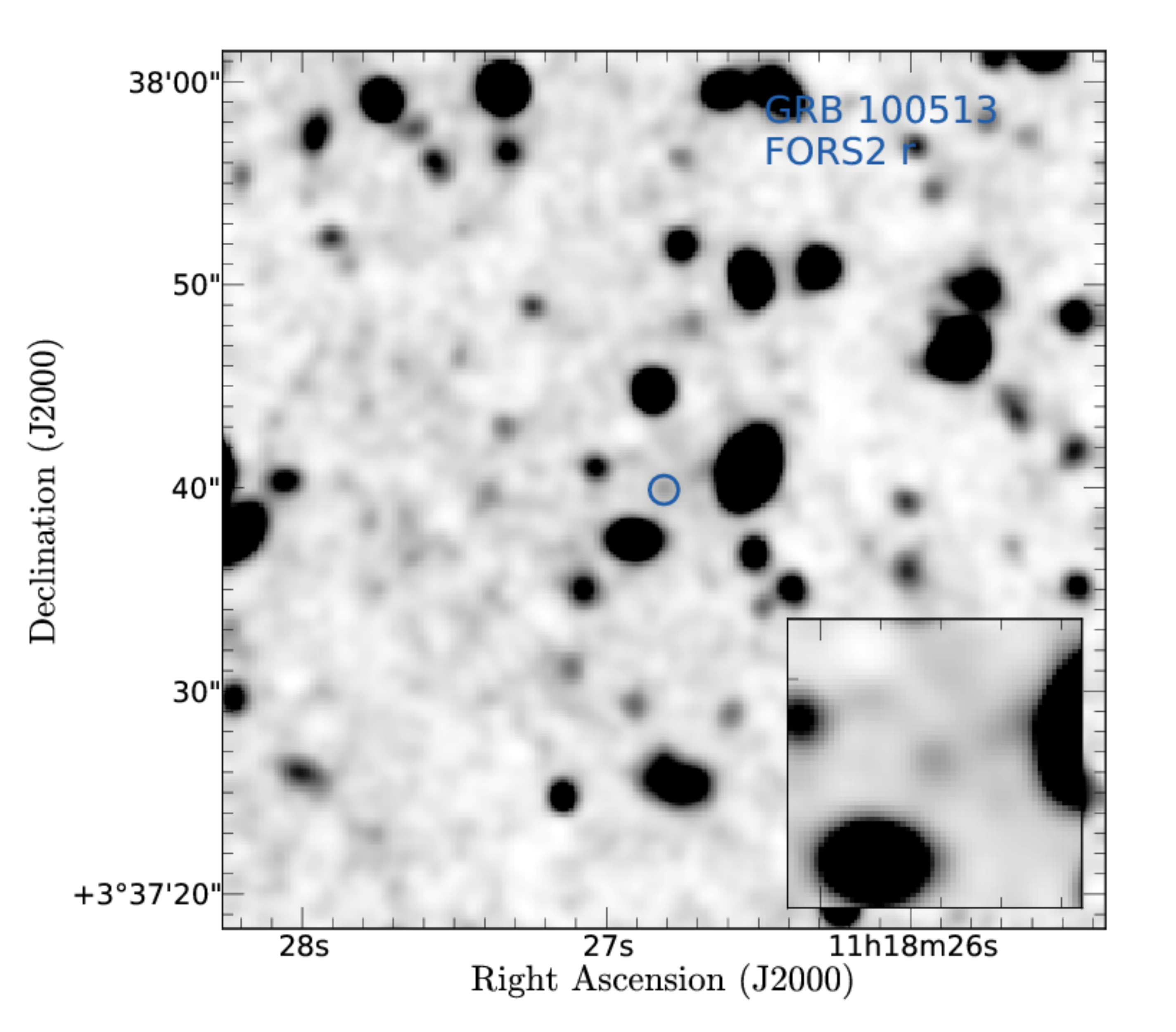} 
\end{figure}

\setcounter{figure}{0}

\begin{figure}[p]
%\ContinuedFloat
\includegraphics[width=7.4cm]{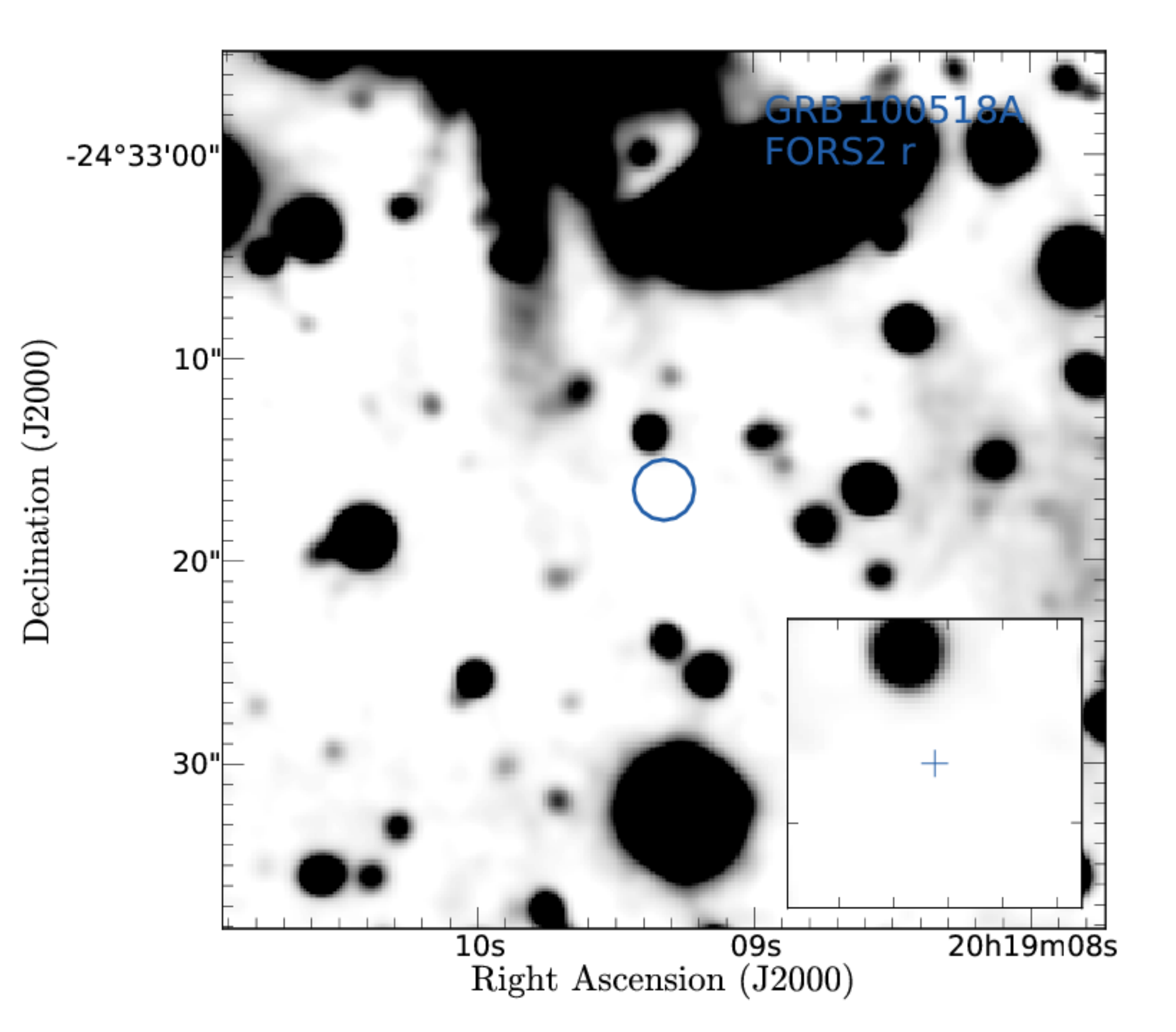} 
\includegraphics[width=7.4cm]{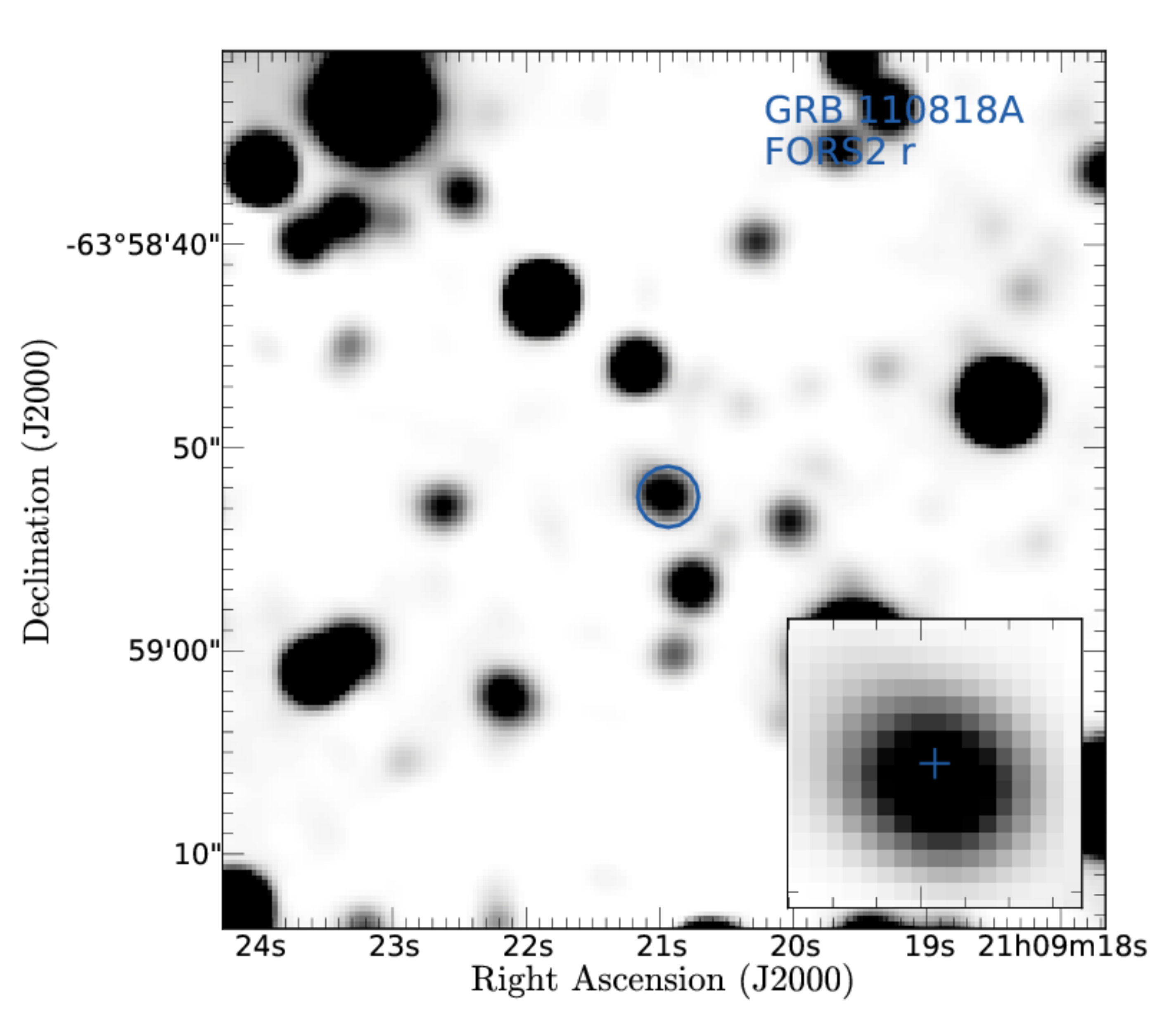} \\
\includegraphics[width=7.4cm]{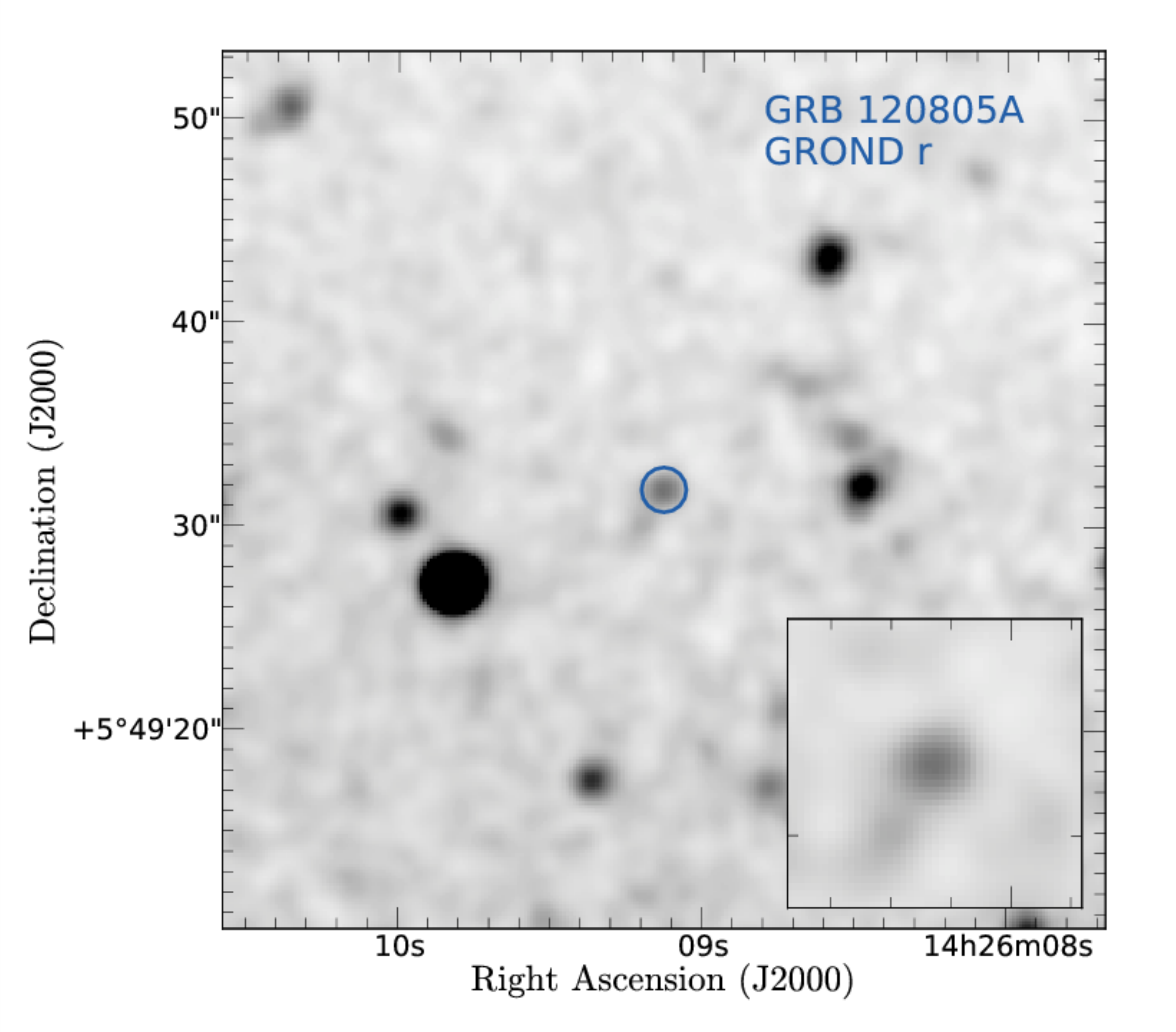} 
\includegraphics[width=7.4cm]{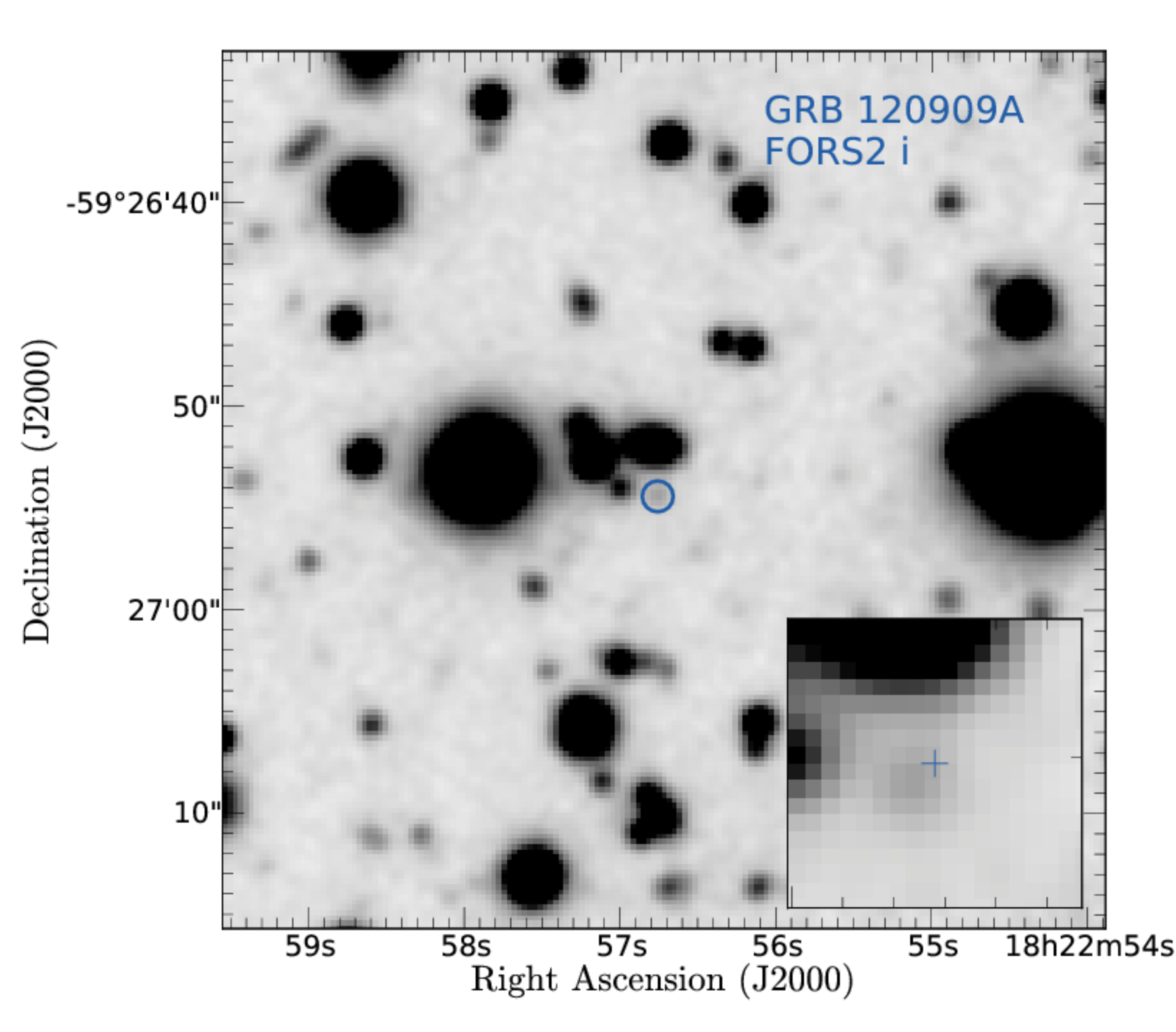} \\ 
\includegraphics[width=7.4cm]{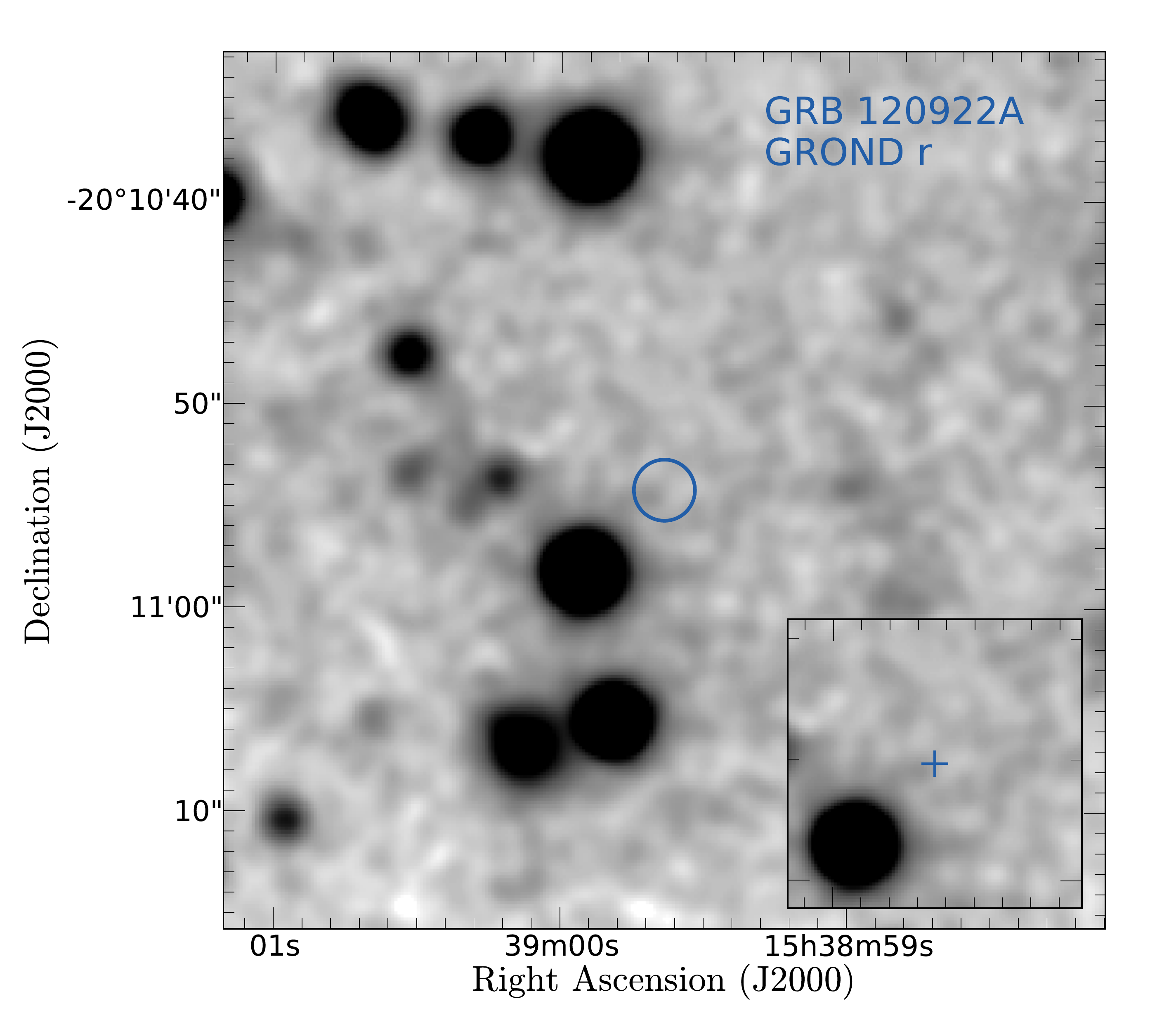} 
\includegraphics[width=7.4cm]{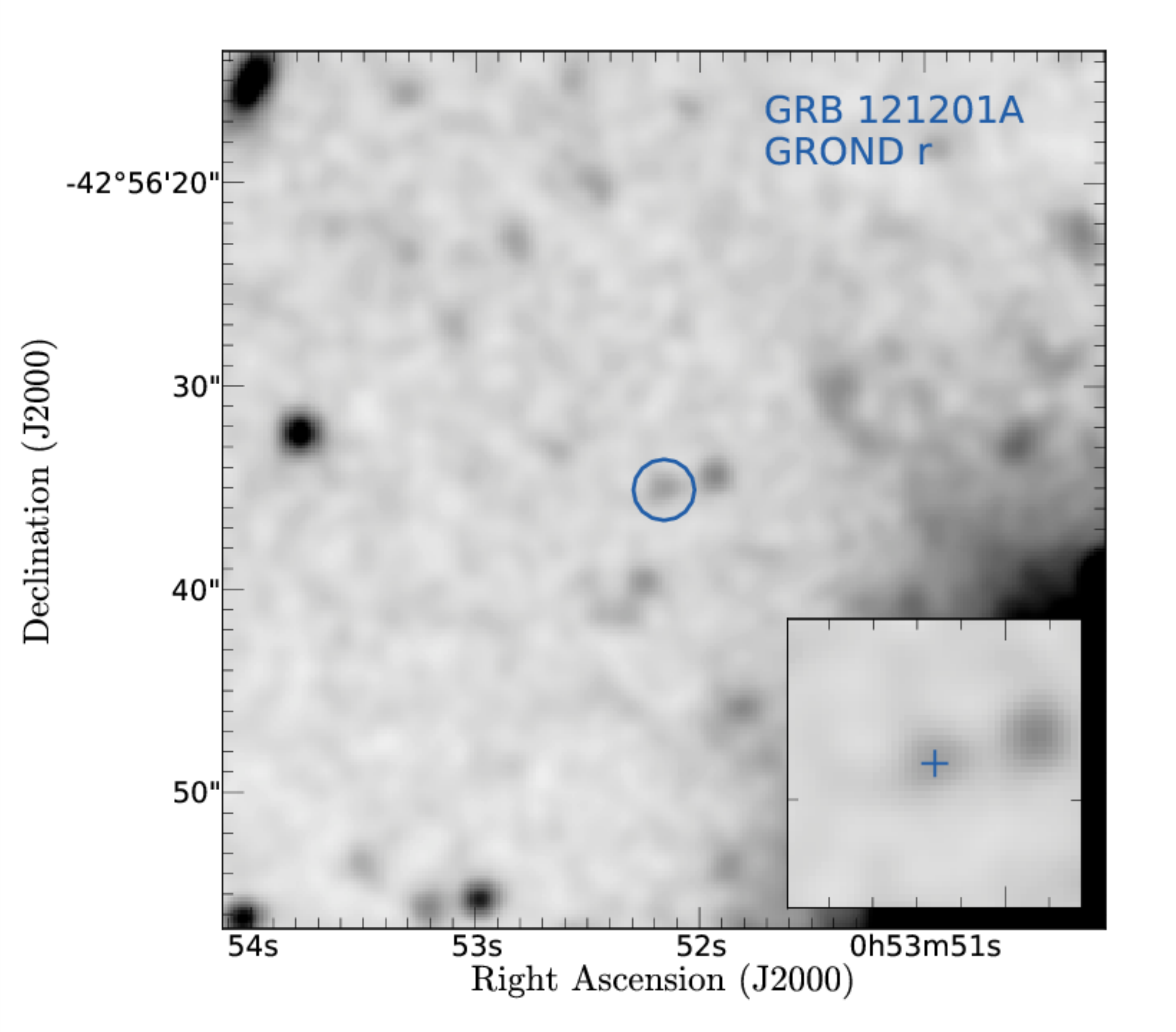} 
\end{figure}

\setcounter{figure}{0}

\begin{figure}[th]
\includegraphics[width=7.4cm]{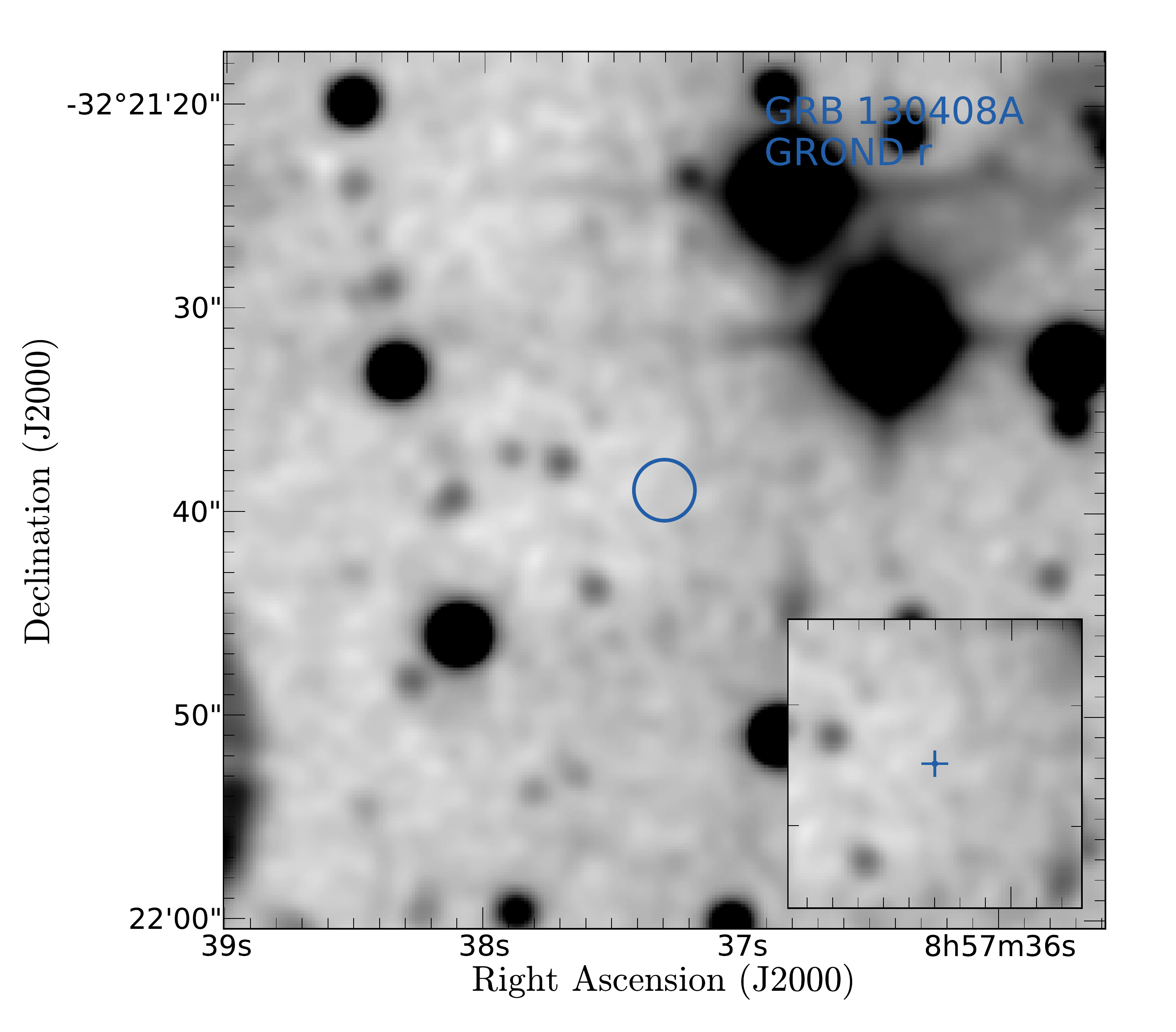}
% 120922A 121201A 130408A
\vspace{-0.3cm}
\caption[]{Finding charts of our new GRB host observations; the GRB name,
instrument and filter band are labeled in each chart. 
North is up, and East to the left.
The circle is centered on the host (if detected) or on the afterglow position,
and denotes the aperture used to extract the photometry. 
%For the GRB with detected host, 
The cross indicates the afterglow position.
\label{fig:obs}}
\end{figure}

%%%%%%%%%%%%%%%%%%%%%%%%%%%%%%%%%%%%%%%%

With respect to the data reduction for individual GRB positions, we
have the following notes:

{\bf GRB 090313:}
The galaxy 2\farcs5 North of the GRB afterglow position is not the
host galaxy. Using Le PHARE\footnote{LePHARE website:
  \url{http://www.cfht.hawaii.edu/$^\sim$arnouts/LEPHARE}}
\citep{Arnouts99, Ilbert06}, the spectral energy distribution obtained
with GROND over all seven optical+NIR channels is best fit with an Sbc
spiral galaxy template at a photometric redshift of $z_{\rm phot} =
1.0 \pm 0.1$.  This is much smaller than the GRB afterglow redshift,
and also lower than the redshifts of the two intervening absorber
systems reported by \citet{deUPostigo09}.  The limiting magnitude of
our GROND image is about 1\fm5 deeper than reported in Table \ref{tab:obs}, 
but the halo of this nearby
galaxy produces additional background light at the afterglow position.

{\bf GRB 120805A:}
Based on the combination of GROND afterglow photometry and X-Shooter
host spectroscopy, \citet{kmf+15} derive a photometric redshift
$z\approx 3.1$.

%%%%%%%%%%%%%%%%%%%%%%%%%%%%%%%%%%%%%%%%

Table~\ref{tab:hosts} presents further data on these observed GRBs,
along with our compilation of GRB host measurements from the
literature. Data columns present the GRB name and redshift (columns 1
and 2), the host magnitude or upper limit in the measured filter band
(3), the Galactic foreground extinction from \cite{schlafly+11} (4),
and the distance modulus for our chosen cosmology (5); the derived
value of the spectral slope $\beta$ (6), absolute UV magnitude
\Mrfuv\ (7), and inferred star formation rate (8); and appropriate
references (9).

%%%%%%%%%%%%%%%%%%%%%%%%%%%%%%%%%%%%%%%%%%%%%%%%%%

\subsection{Host Galaxy Properties}
\label{sub:hostprops}

We derive host galaxy properties (Table~\ref{tab:hosts}) from the
observed magnitudes and upper limits as follows: 

\begin{enumerate}

  \item We correct observed magnitudes for the effects of Galactic
    extinction, applying the extinction law of \citet{ccm89} with $R_V
    = 3.1$, using the dust maps of \citet{schlafly+11} to estimate
    $E(B-V)$ for each line of sight;

  \item We calculate the absolute magnitude at the observed 
    rest-frame wavelength by applying the distance modulus for our
    chosen cosmology ($H_{\rm o}$=70 km/s/Mpc, $\Omega_{\rm M}$=0.3,
    $\Omega_{\rm \Lambda}$=0.7);

  \item We solve jointly for the absolute magnitude of the host galaxy
    at the rest-frame wavelength $\lambda = 1600$\AA, \Mrfuv, and the
    spectral index of its continuum, $\beta$ ($f_\lambda \propto
    \lambda^\beta$), using the relation for star-forming galaxies at
    $\zmean = 3.8$ from Table~3 of \citet{bio+14}:

    \begin{equation}
        \beta = -1.85 - 0.11 (\Mrfuv + 19.5).
    \end{equation}

    \newtext{Note that while this relation exhibits substantial
      scatter, uncertainties for the fitted parameters over ensembles
      of high-redshift star-forming galaxies are modest,}
    $\beta_{-19.5} = -1.85(1)(6)$ and $d\beta/dM_{\rm UV} = -0.11(1)$.

    Our procedure for upper limits is as follows: If our upper limit
    implies $\Mrfuv > -18.1$\,mag (strong constraint) then we adopt
    the $\beta$ value appropriate for a galaxy with absolute magnitude
    equal to our upper limit. If our upper limit is at $\Mrfuv <
    -18.1$\,mag (weak constraint) then we adopt $\beta=-2$
    exactly. Note that for the latter observations, our quoted
    \Mrfuv\ limits and our observed upper limits (at rest-frame
    wavelength $\lambda_{\rm obs}$) will be the same.

  \item In a handful of cases (GRBs 000131, 060223A, 060605, 081029, 090323, 
    090519, 100518A, 130408A) observations of host galaxies at
    $z>3.5$ are attempted in the $R$-band or similar filters, which
    extend over the Lyman-alpha feature at these redshifts. We correct
    for absorption blueward of the Lyman-alpha transition by
    integrating the adopted power-law continuum over the filter
    bandpass, assuming a uniform 60\% suppression of the host galaxy
    continuum at wavelengths blueward of Lyman-alpha. This
    $\approx$1\,mag of UV continuum suppression is typical for
    high-redshift LBGs (e.g., \citealt{aoi+04}). The corresponding
    \Mrfuv\ values are marked with an asterisk in Table
    \ref{tab:hosts}. 

  \item We estimate the host galaxy star formation rate  from the
    calculated spectral slope $\beta$ and absolute magnitude
    \Mrfuv\ by assuming that the UV extinction $A_{1600} = 4.43 ~{\rm mag} +
    1.99\beta$ \citep{mhc99}, and that the total star formation rate
    is proportional to the extinction-corrected UV luminosity (Eq.\ 11
    of \citealt{dcm+14}). For consistency, we quote estimates from
    each single-band observation, even for hosts with observations in
    multiple filters or with previously-published SFR
    estimates. Extinction values are required to be non-negative; only
    the two faintest host limits are affected by this restriction.

\end{enumerate}

We note that since observed spectral indices of high-redshift
star-forming galaxies are close to $\beta\approx -2$ \citep{bio+14},
their restframe UV spectra are nearly flat in $f_\nu$, with AB
magnitudes that are close to constant redward of the Lyman-alpha
transition. Further, we note that \citet{dcm+14} found the adopted
star-formation estimator satisfactorily reproduced the results of full
spectral energy distribution analyses (to within $<$0.1\,mag) for
high-redshift galaxies in CANDELS GOODS South. On the other hand, we
recognize that the SFR estimates for galaxies detected in multiple
bands are not always consistent; GRB\,110818A represents an extreme
case, with \Mrfuv\ and SFR estimates from $R$ and $I$-band
observations that differ by more than $3\sigma$; the spectral slope
between the two observations is $\beta = +0.28$, very different than
the $\beta \approx -1.6$ expected on the basis of the galaxy's
absolute magnitude (although this relation is observed to have
significant scatter, e.g.\ Fig.~7 of \citealt{bio+14}).

Among the six targets with photometric redshift (or similarly low
accuracy) estimates, the error on the derived absolute magnitude
across the likely redshift range is less than 0.3~mag for all but
two. For the two exceptions, GRBs 100518A and 000131, this is due to
the uncertain impact of Lyman-alpha absorption in $R$-band
observations. Since GRB\,000131 has an $I$-band limit which we use for
our host studies, only the single limit for GRB\,100518A (one of 29
detections and 15 limits) is affected in a significant way by
photometric redshift uncertainty.

As a consequence of the observed correlations between \Mrfuv, $\beta$,
and $A_{1600}$ specified above, we find that (except where the
condition of non-negative extinction applies) $A_{1600} = 0.75 - 0.22
(\Mrfuv + 19.5)$ and hence $M_{\rm UV,int} = 1.22\, \Mrfuv +
{\rm constant}$. Thus we expect the intrinsic UV luminosities of the
host galaxies to scale as $L_{\rm UV,int} \propto L_{\rm
  UV}^{1.22}$ for bright galaxies, $\Mrfuv < -16.1$\,mag. In the
complementary regime, $\Mrfuv > -16.1$\,mag, the expected extinction
is zero, and intrinsic and observed UV luminosities will be
equivalent. Hence, since SFR is proportional to $L_{\rm UV,int}$, we
adopt the above SFR weighting to convert from the observed luminosity
function of LBG galaxies in the field to predicted luminosity
functions for GRB host galaxies (see \S\ref{sub:lf}).

%%%%%%%%%%%%%%%%%%%%%%%%%%%%%%%%%%%%%%%%%%%%%%%%%%

\subsection{Comparison to the Literature}
\label{sub:litcomp}

In general, the absolute magnitudes derived here from previously
published magnitudes and limits are in good agreement with those
already published, in particular for filters located close to
rest-frame 1600\AA\ (corresponding to $\lambda_{\rm obs}=6400$\AA\ at
$z=3$ and $\lambda_{\rm obs}=9600$\AA\ at $z=5$). For
longer-wavelength filters, the effect of the luminosity-dependent
slope relation that we have adopted introduces corrections which range
from 0.2 to 0.4~mag.

\newtext{Three GRBs from our sample have published star-formation
  rates:}

\begin{itemize}

  \item GRB\,971214: compared to our SFR of 9.0$\pm$1.3~\Msunpyr,
    \citet{kdr+98} report 5.2~\Msunpyr\ unobscured SFR, and
    \citet{plt+13} report $58.9^{+31.8}_{-8.9}$~\Msunpyr. 

  \item GRB\,080607: compared to our SFR of
    2.1$\pm$1.0~\Msunpyr\ \citet{plt+13} report
    $19.1^{+7.1}_{-4.9}$~\Msunpyr. 

  % 090323: 6 Msun/yr Savaglio+12
  \item GRB\,090323: compared to our SFR of 26.3$\pm$3.9 \Msunpyr,
    \citet{srg+12} estimate 6.4~\Msunpyr\ obscured SFR from the same
    GROND \rp-band imaging. 

\end{itemize}

\newtext{The difference for GRBs 971214 and 080607 is likely due to
  the different method used, i.e. the SED fitting employed by
  \citet{plt+13} For GRB 090323, the difference is nearly solely
  (factor 3.0 out of the total factor 4.1 difference) due to the
  $\beta$-slope dependent extinction-correction which we (but not
  \citealt{srg+12}) have applied; the two approaches are consistent to
  within a factor of two if the same method is used. The case of GRB
  090323 also demonstrates the relatively small residual effect
  (remaining difference of $\approx$50\%) of even a relatively extreme
  spectral slope ($-1.6$) as derived via our approach.}

%%%%%%%%%%%%%%%%%%%%%%%%%%%%%%%%%%%%%%%%%%%%%%%%%%
%%%%%%%%%%%%%%%%%%%%%%%%%%%%%%%%%%%%%%%%%%%%%%%%%%

\section{Analysis \& Interpretation}
\label{sec:interpret}

We now proceed to analyze our dataset of GRB host galaxy properties in
light of current models of galaxy formation and GRB production,
addressing astrophysical biases and possible selection effects. 

%%%%%%%%%%%%%%%%%%%%%%%%%%%%%%%%%%%%%%%%%%%%%%%%%%

\subsection{Host Galaxy Luminosity Function}
\label{sub:lf}

First, we use our collection of $3<z<5$ GRB host galaxy luminosities
and upper limits to estimate the GRB host galaxy luminosity function
and compare to recent survey work on $z>3$ galaxies using deep-
and wide-field imaging observations from HST. For each burst we choose
the single best estimate or upper limit from Table~\ref{tab:hosts},
preferring detections over limits, and higher-precision measurements
(e.g., HST data) over lower-precision measurements, and selecting the
shortest-wavelength filter redward of Lyman-alpha to minimize
systematic uncertainty.

\begin{figure}[thp]
\centerline{~\includegraphics[width=16.0cm]{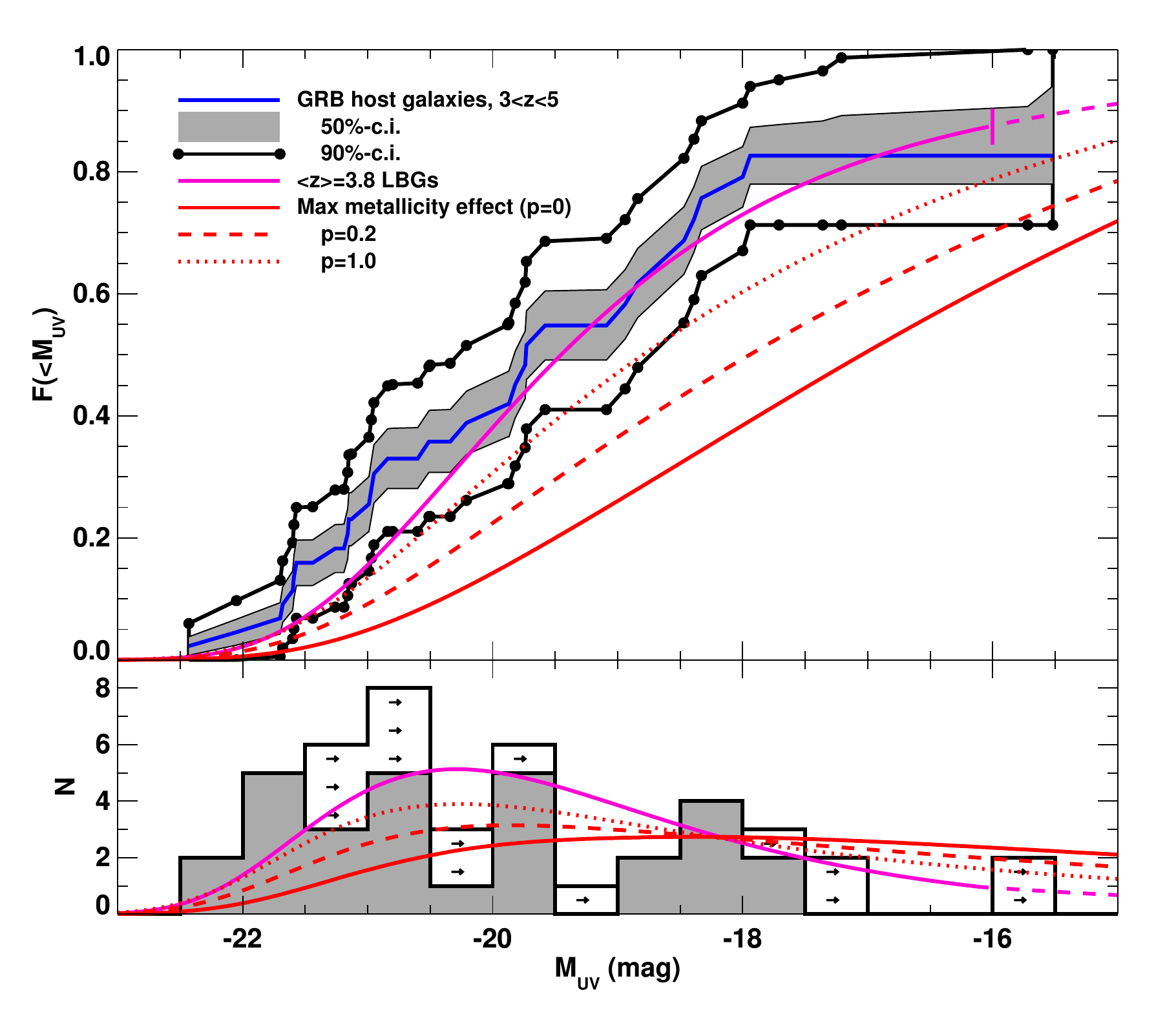}~}
\caption{\footnotesize%
%%% 
  Luminosity function (LF) for GRB host galaxies over $3 < z < 5$ and
  comparison to the Lyman-break galaxy (LBG) LF at $\langle z \rangle
  = 3.8$ \citep{bio+15} and expectations from semi-analytic models
  incorporating anti-metallicity bias in GRB production \citep{tpj15}.
%%%
  Top panel: Cumulative \Mrfuv\ LF for $3<z<5$ GRB host galaxies, as
  derived from our detections and upper limits using Kaplan-Meier
  estimation. The maximum-likelihood luminosity function is plotted as
  a blue line, with the 50\%-confidence region shaded in grey and
  90\%-confidence upper and lower bounds plotted as black lines with
  points. It is compared to the cumulative star formation rate
  (SFR)-weighted LF of $\zmean = 3.8$ LBGs from \citet{bio+15}
  (magenta line), \newtext{and to three predicted $z=3.75$ GRB host galaxy
  luminosity functions from \citet{tpj15} (red lines). The strength of
  anti-metallicity bias in GRB production is characterized by the $p$
  parameter: $p=0$ implies no GRB production at metallicities greater
  than \Zsun, while $p=0.2$ (1.0) implies suppression by a factor of
  six (two) by comparison to the GRB production rate at metallicities
  less than $10^{-3} \Zsun$. The GRB host galaxy LF is consistent with
  the unbiased SFR-weighted LBG LF, borderline-consistent (at
  90\%-confidence) with the $p=1$ model, and inconsistent at $>$90\%
  confidence with the two more biased ($p=0.2$ and $p=0$) models.}
%%%
   Bottom panel: Histogram of $3<z<5$ GRB host galaxy \Mrfuv\ 
   absolute magnitudes and lower limits, as presented in
   Table~\ref{tab:hosts}, along with the SFR-weighted LBG LF at $\zmean =
   3.8$ (magenta line) \newtext{and GRB host galaxy LFs from
   \citet{tpj15}.}  
%%%
\label{fig:kmeier}}
\end{figure}

With the resulting set of 29 \Mrfuv\ measurements and 15 lower bounds,
we use Kaplan-Meier estimation \citep{km58}, as implemented in the
\texttt{R} statistical software package \citep{Rsurvive,Rmanual}, to
construct the maximum-likelihood LF for GRB host galaxies over
$3<z<5$, along with its 50\%-confidence and 90\%-confidence
ranges. Our results are presented in Fig.~\ref{fig:kmeier} (upper
panel), along with a histogram of the input detections and limits
(lower panel). As common in astronomical applications of the
Kaplan-Meier algorithm, the requirement that the ``survival'' of any
individual data point be unrelated to its true value is not strictly
met, since the intrinsically-faintest galaxies are less likely to be
detected. However, as also common in astronomical situations, the
broad range in source redshifts and the range in achieved limits
across the various observations serve to randomize the distribution of
detections and limits somewhat. As such, we are confident that the
resulting characterization of the LF is roughly as good as any
non-parametric analysis of the data can offer (see \citealt{fb12} for
discussion).

For comparison purposes, we use the $\zmean=3.8$ luminosity function
of \citet[][the ``All Fields'' fit from Table~6]{bio+15}. This
luminosity function results from their $3.3\simlt z \simlt 4.5$
photometric redshift selection of Lyman-break galaxies (LBGs) over the
XDF, HUDF09-Ps, CANDELS (North + South), and ERS HST survey fields,
and reaches to $\Mrfuv = -16$\,mag. A Schechter-function form
\citep{schech76} is assumed, and the best-fit parameters with
uncertainties are: \newtext{$M^* = -20.88(8)$\,mag, $\phi^* = 1.97(31)\times
10^{-3}$\,Mpc$^{-3}$, and $\alpha=-1.64(4)$}%
\footnote{\newtext{We note that the best-fit LBG LF parameters changed
    between initial preprints of \citet{bio+15} and the published
    version, owing to an improved treatment of the survey selection
    volume; the original submitted version of the present manuscript
    used the initial parameter values.}}.
In this context, we note that the median and mean redshifts of our
sample are 3.57 and 3.72, respectively.

To compare the LBG and GRB host galaxy luminosity functions, we must
account for GRB selection effects. Our default hypothesis, consistent
with the collapsar model, is to weight galaxies according to their
instantaneous star-formation rates. Here we use results established
previously (\S\ref{sub:hostprops}), showing that the star
formation rate scales as ${\rm SFR} \propto L_{\rm UV}^{1.22}$ for
$\Mrfuv \le -16.1$\,mag, and as ${\rm SFR} \propto L_{\rm UV}$ for
$\Mrfuv > -16.1$\,mag. The effect of an SFR-weighted selection of
galaxies on the luminosity function is thus to increase the $\alpha$
parameter by 1.22 (1.00) to $\alpha_{\rm SFR} = -0.42$ ($-0.64$) over
these two regimes, without altering $M^*$. Note that $\phi^*$ is
irrelevant in the present context as our sample is fixed at $N=44$
galaxies.

Overplotting the cumulative SFR-weighted LBG LF from \citet{bio+15}
(magenta line) against our cumulative LF for the GRB host galaxies
(blue line) shows that the two are consistent across the full range of
absolute magnitudes probed (Fig.~\ref{fig:kmeier}). \newtext{In
  particular, the LBG LF stays within the 90\%-confidence bounds
  (black lines with points) on the GRB host galaxy LF for 38 of 42
  plotted points (exceeding the constraint for four points at $\Mrfuv
  < -20.5$\,mag), and lies within the 50\%-confidence interval (shaded
  region) for 33\% of this range. Exploring the degree of agreement
  quantitatively, we find that the LBG LF is consistent with our
  observations at roughly the 1$\sigma$ (68\%-confidence) level.}

Given consistency with this straightforward model, we do not find
evidence for complex or second-order selection effects for GRB host
galaxies in this redshift range. Effects that have been proposed in
this context include metallicity-dependent selection (biasing toward
lower-mass galaxies; \citealt{fls+06,mkk+08,wd14}) and selection
against the most luminous, dusty hosts (via extinction of the optical
afterglow light; \citealt{pcb+09, kgs+11, plt+13}).  
By contrast, our results are consistent with predictions that
metallicity or anti-extinction effects in the GRB population will
yield minimal selection bias for GRB host galaxies at $z>3$
\citep{kwm09,msc11,plt+13}. \newtext{We explore the question of
  anti-metallicity bias quantitatively in Sect.~\ref{sub:models} below,
  comparing our host galaxy LF to expectations under a range of
  scenarios (their ``plateau parameter'' $p=0$, 0.2 and 1.0) for
  anti-metallicity bias explored by \citet{tpj15}. These models are
  plotted as red lines in Fig.~\ref{fig:kmeier}.}

We note that our results provide independent evidence for host
luminosity-dependent obscuration as observed both at $z\approx 3$
\citep{mhc99} and in higher-redshift LBG studies
\citep{bio+14,dcm+14}. In particular, if we assume that SFR is
proportional to observed UV luminosity $L_{\rm UV}$ across the full
range of galaxy luminosities, the resulting LF is inconsistent with
our GRB host galaxy LF at $>$90\%-confidence, lying to the right of
the illustrated 90\%-confidence interval for 35 of 42 plotted points
\newtext{(and between the plotted LFs for the $p=0.2$ and $p=1.0$ models from
\citealt{tpj15}).}

{\it Comparison to recent work --}
\newtext{Since submission of this manuscript, a separate study of GRB
  host galaxy luminosity functions over $0<z<4.5$, including
  comparison to expectations from galaxy surveys and semianalytic
  models, has been posted to the ArXiv \citep{sch+15}. These authors
  study the 69 GRB positions defined by the TOUGH sample
  \citep{hmj+12}, a subset of which are included in our
  compilation. Since the authors do not perform a Kaplan-Meier
  analysis, the effects of host galaxy nondetections (limits) on their
  LFs are not clear, and they restrict their analysis to the depth of
  the faintest detected host galaxy in each redshift interval,
  foregoing the opportunity to test their LFs from the faint
  end. Nonetheless, they compare their resulting empirical LFs to LBG
  LFs and the semi-analytic LFs from \citet{tpj15}, as we have (see in
  particular their Fig.~7).}

\newtext{\citet{sch+15} find evidence for significant anti-metallicity
  bias over $0 < z < 1$ ($n=13$ galaxies) and $3.0\le z \le 4.5$
  ($n=9$ galaxies). Their empirical LF over $1.0\le z\le 1.5$ ($n=7$
  galaxies) is consistent with all tested models, while the LF over
  $1.9\le z\le 2.7$ ($n=22$ galaxies) disfavors models with strong
  anti-metallicity bias ($p=0$ and $p=0.04$). Over the two
  higher-redshift intervals, their results are subject to significant
  uncertainty due to the unknown redshifts of nine of their targeted
  GRBs, six of which have detected host galaxies. Including a
  significant number of these host galaxies in the $1.9\le z\le 2.7$
  redshift interval would shift the LF for this interval to fainter
  magnitudes, increasing agreement with moderate and strongly biased
  models, while including a significant number in the $3.0\le z\le
  4.5$ interval (owing to the greater implied host galaxy luminosities
  at these redshifts) would shift the LF for this interval to brighter
  magnitudes, increasing agreement with the less-biased models
  (including simple SFR-weighting or no bias, $p\rightarrow\infty$).} 

\newtext{Given the uncertainties, distinct statistical approach, and
  relatively small sample size within the redshift range considered
  here, we do not consider the results of \citet{sch+15} to be in
  significant tension with ours. To the contrary, we think they
  demonstrate (along with our present work) that GRB host galaxy
  studies have advanced to the point where LF comparisons over a broad
  range of redshifts are providing useful insights into the host
  galaxy population, and likely, the GRB mechanism itself.  }

%%%%%%%%%%%%%%%%%%%%%%%%%%%%%%%%%%%%%%%%%%%%%%%%%%%%%%%%%%%%

\subsection{Astrophysical Effects}
\label{sub:models}

The consistency we have demonstrated between the GRB host galaxy LF
over $3<z<5$ and the most straightforward predictive model -- the
SFR-weighted LF of $\zmean=3.8$ LBGs from \citet{bio+15} -- suggests
that astrophysical biases in selecting GRB host galaxies at these
redshifts are either minor or counterbalanced by competing
astrophysical or selection effects.

Of course, it may be the case that these astrophysical biases are,
indeed, present and counterbalanced. Even if not, reviewing past and
current predictions as to the nature and strength of these effects,
and comparing to the constraints implied by our data, may help to
place GRBs and their host galaxies in their full cosmic context. We
therefore review these questions here, considering first proposed
astrophysical biases and then the most likely observational or
selection effects.

% Metallicity

{\it Metallicity bias:}
The effects of host galaxy metallicity on GRB production have been
much debated. From the standpoint of the collapsar model of
long-duration GRBs \citep{woosley93,mfw99}, the metal content of a
massive star's envelope is a strong driver of mass and angular
momentum losses. Expulsion of the hydrogen envelope before core
collapse is necessary to reduce the star's size to the point where the
GRB jet can successfully escape the star while the central engine is
active. At the same time, if angular momentum losses from the envelope
are transmitted to the core, they will prevent a long-lived accretion
disk from forming, drastically reducing the lifetime of the central
engine. As a result, isolated high-metallicity stars are not expected
to explode as GRBs \citep{wh06,yln06}. The strength of this bias could
be mitigated, however, if binary stellar evolution can produce GRB
progenitors more or less independent of metallicity
\citep{fh05,cyl+07}.

From an observational perspective, a bias toward low-metallicity
environments finds support from the lowest-redshift ``low luminosity''
GRBs, which have been found exclusively in low-metallicity host
galaxies and regions of those galaxies \citep{sgb+06,lbs+11}; from
metallicity studies of GRB host galaxies at $z\simlt 1$
\citep{lkb+10,gf13} and beyond \citep{kmf+15}; from comparative host
galaxy studies of GRBs, type Ibc, and type II supernovae at $z\simlt
2$ \citep{fls+06,mkk+08}; and from GRB host galaxy photometric surveys
at $z\simlt 2$ \citep{plt+13} and beyond
\citep{fjm+03,hmj+12}. However, contrary indications also exist: for
example, not all GRB host galaxies are low-mass or metal poor
\citep{sglb09,lkg+10,srg+12,ekg+13,skg+15,kmf+15}, and gas-phase
metallicities derived from GRB absorption spectroscopy are consistent
with average galaxy metallicities for star-forming galaxies at similar
redshifts \citep{pcd+07,cfr+14}. Moreover, with respect to
low-luminosity GRBs it should be noted that these are not, strictly
speaking, GRBs -- their much lower luminosities can be accommodated
within mildly-relativistic shock-breakout models, without the need for
highly-relativistic jets \citep{ns12,ksw12,bdnps14}.

In an attempt to shed light on these issues, \citet{tpj15} extended
their semi-analytic models of galaxy formation to incorporate
metallicity-sensitive GRB production rates.  They explored a range of
models, from zero to near-maximal sensitivity, by including
metallicity-sensitive (MS) and metallicity-insensitive (MI) channels
of GRB production in a variable ratio. The redshift distribution and
host galaxy properties of the GRBs in each scenario are then compared
to observations. Importantly, with these models they address, for the
first time and in realistic fashion, the effect that the evolving
metallicity distribution of star-forming regions, in galaxies of
various sizes, will have on the metallicities, masses, and
luminosities of GRB host galaxies across the full range of observed
redshifts.

\citet{tpj15} find that the \swift\ GRB redshift distribution
\citep{wp10} is best reproduced by models that incorporate
contributions from both MS and MI production channels. In their
maximum-likelihood scenario, most GRBs at $z\simlt 1$ result from the
MI channel and hence reflect the typical metallicities of star-forming
regions; GRBs are produced at comparable rates in the two channels at
$z\approx 3$ and show the strongest differential preference (compared
to total star-formation metrics) for low-metallicity regions; and GRBs
are produced primarily via the MS channel at $z\simgt 5$, when nearly
all star-formation is occurring in metal-poor environments.

% possible comment on metallicity-aversion countering effects of dust
% obscuration on SFR estimation from L_UV here \fix

% could also point out the distinct approaches to SFR ~ L_UV^x scaling
% \fix 

We have compared our host galaxy LF to the LF predictions of
\citet{tpj15} at $z=3.75$ (Fig.~\ref{fig:kmeier}). Our LF is
consistent with their predictions for the pure-MI scenario and
inconsistent at $>$90\%-confidence with the pure-MS scenario (with
their ``plateau'' variable $p=0$), as well as with two mixed scenarios
($p=0.04$, not plotted, and $p=0.2$) that incorporate both MS and MI
GRB production. \newtext{It is marginally consistent (not excluded at
$>$90\%-confidence) with a further $p=1$ mixed scenario that was
generated specifically for this work. }
The plateau variable sets the relative efficiency of GRB production
via the MI channel at metallicities $Z\ge \Zsun$: GRB production is
suppressed by a factor of $(1+p)/p$ (e.g., by 6$\times$ for $p=0.2$,
and by 2$\times$ for $p=1$) for $Z\ge \Zsun$, compared to its rate at
$Z\le 10^{-3}\,\Zsun$. In a pure-MS scenario, $p=0$ and there is no
GRB production at $Z\ge \Zsun$, while as $p\rightarrow \infty$ GRB
production is independent of metallicity. In models with finite $p$,
the GRB production efficiency at intermediate metallicities $10^{-3} <
Z/\Zsun < 1$ decreases monotonically with increasing metallicity; see
\citet{tpj15} for details.

Consistency with the pure-MI model is expected, as it is constructed
in part on the LBG survey work of \citet{bio+15} which provides our
successful default LBG LF (\S\ref{sub:lf}). The incompatibility of the
$p=0.2$ mixed model with our observed LF is more surprising, as it
successfully reproduces the GRB redshift distribution. Yet our sample
includes a significantly greater proportion of bright host galaxies
than this model predicts, \newtext{and is barely consistent with the
  significantly reduced bias of the $p=1$ mixed model.} We conclude
that a dominant role for the MS channel in GRB production at $3<z<5$
can only be supported if some distinct astrophysical or selection
effect is acting to enrich our sample with bright host galaxies.

%%%%%%%%%%%%%%%%%%%%

% Msupp

{\it Suppression of star formation in low-mass halos:}
In fact, the possible existence and likely implications of one such
astrophysical effect have been central to several earlier studies
relating GRB production to high-redshift star formation
\citep{tpl+12,tpt13}. These models assume that SF at $z>3$ is
efficient in dark matter halos only above a certain minimum mass,
leading to a Schechter-like LF up to a corresponding maximum
(faintest) \Mrfuv, beyond which the LF cuts off sharply. This limiting
``suppression magnitude'' \Msupp\ has been taken to be $-12$\,mag in
the so-called standard scenario, corresponding to a minimum halo mass
of $10^8$\,\Msun.  A significantly larger minimum halo mass for
successful star formation would bias the LF to brighter magnitudes and
increase the fraction of detected host galaxies at a given flux
limit. Hence, the allowed range for \Msupp\ can be constrained using
observations of host galaxies at substantially brighter magnitudes,
$\Mrfuv \ll \Msupp$.

Such an analysis has previously been applied by \citet{tpl+12} to the
deep limits for six $z>5$ GRB host galaxies reported by
\citet{tlf+12}; they showed that the luminosity function at those
redshifts must extend to at least $\Msupp > -15$\,mag (90\%-confidence
limit). With respect to the present dataset and our lower-redshift
range, $3<z<5$, the non-detection of two host galaxies down to $\Mrfuv
> -15.52$\,mag (GRB\,060607A at $z=3.075$) and $\Mrfuv > -15.72$\,mag
(GRB\,020124 at $z=3.198$) provides a strong lower bound, $\Msupp >
-15.52$\,mag, that is almost as constraining as the previous $z>5$
result.

Because a non-negligible fraction of SF happens in faint galaxies for
any Schechter-like LF, values of \Msupp\ that are consistent with our
lower (bright-end) bound can still alter the predicted GRB host galaxy
LF in a significant fashion. For example, the $p=0.2$ LF for $z=3.75$
galaxies from \citet{tpj15} is consistent with (not ruled out at
$>$90\%-confidence by) our results for $\Msupp\le -15.6$\,mag. While
this particular case is excluded (in a distinct manner) by our
faintest limits, \newtext{plateau parameter values $0.2 < p < 1$,
  excluded at $>$90\%-confidence for $\Msupp=-12$\,mag, would be
  consistent with our data for values of $\Msupp > -15.5$\,mag which
  are allowed by our deepest limits.}

%%%%%%%%%%%%%%%%%%%%%%%%%%%%%%%%%%%%%%%%%%%%%%%%%%

\subsection{Potential Observational Biases}
\label{sub:biases}

We now turn to consider potential biases in the observations,
independent of astrophysical effects, which may affect our results. We
have identified three such potential biases, and address each of these
in turn.

%%%%%%%%%%%%%%%%%%%%%%%%%%%%%%

{\it Interloper galaxies:}
Some of the identified galaxies, while coincident with the GRB
positions, may not actually be the GRB host galaxies; we refer to this
as the interloper problem. When interlopers are present, the sample
will be contaminated with galaxies having greater inferred luminosity
than the (unobserved) true host galaxies. The nature of this effect,
then, is to bias the observed LF to brighter magnitudes.

We estimate the magnitude of this effect by reference to $R$-band
galaxy number counts from \citet{Boutsia+2014}. We find that for a
$1\arcsec$ radius localization, there is a 1\% chance for an
interloper galaxy with $R_{\rm AB}$ magnitude within 0.25\,mag of
$R_{\rm AB} = 25$\,mag; a 2\% chance for an interloper galaxy within
0.25\,mag of 26\,mag; a 5\% chance for an interloper galaxy within
0.25\,mag of 27\,mag; and a 14\% chance for an interloper galaxy
within 0.25\,mag of 28\,mag. By its nature, the interloper effect is
not relevant for (cannot affect) our upper limits.

Our faintest detection is $m_{\rm 775w} = 28.07(20)$\,mag for the
candidate host galaxy of GRB\,060605; however, this galaxy has a
brighter $R$-band detection from the ground at $R=26.4(3)$\,mag. The
next two faintest detections are $m_{\rm 110w}=27.96(15)$\,mag for
GRB\,060223A and $R=27.76(45)$\,mag for GRB\,050908. These two
detections should be compared to the $\approx$12\% chance of an
interloper of similar magnitude; on a purely probabilistic basis there
is a 78\% chance that both are true associations and a 22\% chance
that one or both are false. Our brighter detections are less likely to
be interlopers; the estimated total number of interlopers across our
full sample of 29 detected host galaxies is 0.85.

If one or more of our detected galaxies are interlopers, identifying
and correcting for this would convert one, or several, of the fainter
detections (e.g., those identified above), into upper limits. Given
the sample size and the estimated number of interlopers, we do not
expect this to have a dramatic impact on our derived LF.

Moreover, we note that we have calculated interloper probabilities
from a frequentist perspective, without taking into account the prior
expectation that a significant fraction of GRB positions should have
star-forming galaxies that will be detected via the reported
observations. A full Bayesian analysis, beyond the scope of the
present paper, would be expected to yield a smaller contamination rate
than we have calculated above, with a correspondingly weaker effect on
the derived LF.

%%%%%%%%%%%%%%%%%%%%%%%%%%%%%%

{\it Dark burst host galaxies:}
Since we require a spectroscopic or photometric redshift to include a
GRB in our sample, GRBs that experience high extinction in their host
galaxies are more likely to be excluded. The known hosts of confirmed
high-extinction ($A_V > 1$\,mag) GRBs, almost all at $z < 3$, have
been shown to be statistically more massive, more luminous, and more
evolved than the host galaxies of low-extinction bursts in this
redshift range \citep{kgs+11,plt+13}. LBG surveys also uniformly find
that more massive and luminous galaxies are subject to greater amounts
of dust obscuration than less massive galaxies \citep{dcm+14,bio+15}.
Hence, selection against optically-extinguished bursts may exclude
some of the most luminous GRB host galaxies, and bias our derived LF
toward fainter luminosities. \newtext{In the present context, with an
  empirical LF that lies on the higher-luminosity side of the
  SFR-weighted LBG LF and all semi-analytic LFs incorporating effects
  of anti-metallicity bias (Fig.~\ref{fig:kmeier}), detailed
  consideration of this correction may appear
  unwarranted. Nonetheless, our concern with all astrophysical and
  selection effects leads us to review this issue briefly here.}

The most ambitious survey of high-extinction host galaxies to date
\citep{plt+13} -- which first clarified the extent to which $z<3$
hosts of high-extinction bursts are systematically more massive and
luminous than those of low-extinction bursts -- shows a clear decrease
in this trend with redshift (e.g., in their Fig.~7), with the
distributions of high-extinction and low-extinction host galaxies
becoming consistent with each other and with the distribution for
star-forming galaxies by $z\approx 2$ (see caption to their Fig.~8).
Unless this trend reverses at higher redshift, this suggests that the
host galaxies of high-extinction and low-extinction bursts will not be
strongly distinguished over our redshift range of
$3<z<5$. Furthermore, the host galaxies of heavily dust-extinguished
GRBs are not exclusively massive, with 50\% of the sample included in
\citet{plt+13} having stellar masses $M_\star\simlt 10^{10}\,\Msun$,
and less than 20\% having $M_\star\simgt 10^{11}\,\Msun$.

We attempt to quantify the number of host galaxies missing from our
sample due to high afterglow extinction (defined as $A_V>1$\,mag in
the host galaxy) by considering two GRB samples that are relatively
complete in terms of both redshift and $A_V$ measurements. These are
the GROND 4-hour sample \citep{gkk+11}, consisting of 39 long-duration
GRBs observed by GROND within 4 hours of the \Swift\ trigger, and the
BAT peak flux-selected sample \citep{scv+12, cms+13}, made up of 58
GRBs, 11 of which are in the GROND 4-hour sample. Removing duplicates,
the union of these datasets provides a sample of 86 GRBs, 78 of which
have redshift measurements (91\% completeness). Ten GRBs (13\%) have
$3<z<5$, and one of these has $A_V>1$\,mag (GRB\,080607 with
$A_V=1.3$\,mag; \citealt{pmu+11}).  This gives us a lower limit of
10\% on the fraction of GRBs at $3<z<5$ with $A_V>1$\,mag. If we
assume that all eight bursts without measured redshifts are
high-extinction bursts, with redshifts distributed similarly as for
the bursts with measured redshifts, then one of these high-extinction
bursts with unknown redshifts (and two or fewer, at 90\%-confidence)
are from $3<z<5$. This would imply an expected fraction of 18\% of
GRBs ($<$25\%, at 90\%-confidence) over $3<z<5$ with $A_V>1$\,mag.

Since two high-extinction bursts are included in our sample
(GRBs\,080607, \citealt{plt+13}; and GRB\,060210, \citealt{ckh+09}), a
high-extinction fraction of 18\% would mean that we have missed
$\approx$7 high-extinction afterglows in our sample of 44 GRBs. If, as
we expect, the host properties of high-extinction bursts in our
redshift range are not strongly distinguished from those of
low-extinction bursts, then the absence of these host galaxies from
our sample will not significantly affect our results. Alternatively,
if the hosts of high-extinction and low-extinction bursts are strongly
distinguished over $3<z<5$, then our LF has been biased to lower
luminosities by this effect, \newtext{and any correction would shift
  our LF toward even higher luminosities.}

%%%%%%%%%%%%%%%%%%%%%%%%%%%%%%

{\it Publication bias:}
The majority of our host measurements are drawn from the published
literature, including a diverse array of efforts and facilities. This
is advantageous in the sense that it provides coverage of many GRB
positions, yields detections of multiple bright host galaxies, and
provides a somewhat-randomized distribution of detections and upper
limits, helping to make a Kaplan-Meier approach to LF construction
more appropriate. On the other hand, it means the sample may be
affected by publication bias: if upper limits from moderate-depth
observations are considered less interesting, they may be less likely
to be reported in the literature than detections. As a result, our
sample could be depleted in moderate-depth upper limits by comparison
to a hypothetical survey offering complete coverage to a uniform
depth; absence of these upper limits would bias our derived LF toward
brighter magnitudes.

While we cannot guarantee that this effect is absent in our sample, we
have taken care to search the Gemini, VLT, and HST archives for
unpublished late-time imaging of GRB positions within our targeted
redshift range; this process led (in part) to the present set of
coauthors, and to our presentation of five previously-unpublished HST
observations from \citet{svth11}. Moreover, our sample includes
significant contributions from the pre-defined and complete survey
efforts of \citet{hmj+12} and \citet{plt+13}. We thus expect the
impact of future observations of GRBs within the targeted redshift
range, with respect to this effect, to be modest.

%%%%%%%%%%%%%%%%%%%%%%%%%%%%%%%%%%%%%%%%%%%%%%%%%%

\section{Conclusions}
\label{sec:conclude}

We have presented deep ground-based imaging of thirteen GRBs over
$3<z<5$ (Fig.~\ref{fig:obs}), yielding discovery of eight new GRB host
galaxies along with five deep upper limits
(Table~\ref{tab:obs}). Combining these results with published
observations of 31 additional GRB positions, including 21 detected
host galaxies, we have presented a comprehensive and uniform summary
of the photometric properties of GRB host galaxies over $3<z<5$
(Table~\ref{tab:hosts}).

% Results of LF analysis

Adopting ultraviolet (UV) continuum spectral index ($\beta$) and
extinction ($A_{1600}$) prescriptions for high-redshift star-forming
galaxies from multiband Lyman-break galaxy surveys with the
Hubble Space Telescope \citep[HST;][]{mhc99,bio+14,dcm+14,bio+15}, we
have used our detections and limits to estimate host galaxy absolute
magnitudes at $\lambda=1600$\AA\ in the rest frame, \Mrfuv\ (quoted in
AB mags), and their resulting inferred star formation rates. 
Adopting the single most useful detection or upper limit for
each of 44 targeted host galaxies, we have constructed the luminosity
function (LF) for GRB host galaxies over $3<z<5$ by Kaplan-Meier
estimation, calculating the maximum-likelihood LF as well as its
50\%-confidence and 90\%-confidence intervals (Fig.~\ref{fig:kmeier}).

We have compared the GRB LF to expectations from LBG surveys by
constructing the SFR-weighted LF for $\zmean=3.8$ LBGs
\citep{bio+15}. \newtext{This LBG LF has $M^* = -20.88(8)$\,mag,
  $\phi^* = 1.97(31)\times 10^{-3}$\,Mpc$^{-3}$, and
  $\alpha=-1.64(4)$;} our adopted SFR-weighting, appropriate if GRBs
trace star formation, adjusts the power-law slope by
$\Delta\alpha=+1.22$ (+1.00) over $\Mrfuv < -16.1$\,mag ($\Mrfuv
>-16.1$\,mag) owing to the luminosity-dependent extinction observed in
LBG surveys \citep{mhc99,dcm+14}. The resulting LF is
\newtext{compatible} with our GRB host galaxy LF over the full range
of host galaxy luminosities probed, from $\Mrfuv = -22.5$\,mag to
$\Mrfuv > -15.6$\,mag, a range of more than 600$\times$ in host
luminosity. Since GRB host galaxies are selected independent of host
luminosity, our results demonstrate consistency of the two LFs -- with
respect to their Schechter-function form and $M^*$ and $\alpha$
parameters -- well beyond the range of absolute magnitudes that has
been probed directly.

% Astrophysical & Observational Biases

We have reviewed proposed astrophysical and observational effects that
might bias our dataset and resulting LF. If two or more of these
effects are present in sufficient and counter-balancing strength, they
might conspire to yield a misleading agreement between the GRB host
galaxy LF and the SFR-weighted LF of the LBG population, which is
meant to reflect UV metrics of star formation over this redshift
range. 

% Metallicity

In this context, we reviewed the issue of the proposed
anti-metallicity bias of GRBs. Existing observational evidence of this
bias has been gathered almost entirely at low-redshift, $z\simlt 3$,
via studies of GRB host galaxies \citep[][and references
  therein]{levesque14}, and by comparison of the redshift evolution of
the GRB rate to other metrics of star formation
\citep{kyb+08,kyb+09,wp10,re12}. At $z\simgt 3$ a much greater
fraction of star formation occurs in low-metallicity environments,
where the effects of such a bias might be reduced. Nonetheless, recent
theoretical explorations demonstrate that, for plausible models of the
bias, the GRB host galaxy LF can be significantly altered over the
redshift range considered here \citep{tpj15}. We tested these models
against our observed LF and found (in the absence of other biases) no
evidence for a \newtext{dominant} contribution from any
metallicity-sensitive GRB production channel over $3<z<5$.

% M_supp models

Alternatively, suppression of star formation in low-mass halos could
result in a brighter GRB host galaxy LF than predicted via naive
extrapolation of a Schechter-type LF \citep{tsb+10,tpt13}. We do not
find it necessary to invoke this effect; the two faintest host galaxy
limits in our sample, $\Mrfuv > -15.72$\,mag and $\Mrfuv >
-15.52$\,mag, imply directly that $\Msupp \simgt -15.5$\,mag, and are
almost as constraining as those derived from deep HST observations of
known $z>6$ GRB positions \citep{tlf+12,tpl+12}. It remains possible
that this effect is present and compensating for the effects of a
(relatively mild) anti-metallicity bias in the GRB host galaxy
population.

%%%%%%%%%%%%%%%%%%%%%%%%%%%%%%

% Interloper galaxies

In terms of observational or selection effects, we have reviewed our
sample for possible contamination by line-of-sight interloper
galaxies, unrelated to the GRB, and concluded that there is some
chance that one or a few of our faintest detections are spurious in
this sense; we estimated that on average, 0.85 of the host galaxies in
the present sample will be interlopers.

% Dark bursts

The observational challenges of measuring spectroscopic or photometric
redshifts for high-extinction ($A_V > 1$\,mag in the host galaxy) GRB
afterglows are real and likely to affect our sample; we estimate that
we are missing $\approx$7 GRBs as a result of these effects. If the
properties of the host galaxies of these ``dark bursts'' are
significantly different than those of the $A_V < 1$\,mag population,
this could affect our LF and conclusions. However, the most extensive
existing survey of the host galaxies of high-extinction bursts has
shown that the differences between the high-extinction and
low-extinction host galaxy populations decreases with redshift,
becoming quite modest by $z\approx 2$ \citep{plt+13}.

% Publication bias

We have also considered whether our sample might be depleted in
medium-sensitivity upper limits on GRB host galaxies owing to
publication bias. Given our review of the Gemini, VLT, and HST
archives, we consider this possibility unlikely, although it is
difficult to rule out completely.

%%%%%%%%%%%%%%%%%%%%%%%%%%%%%%

Rather than invoke a hypothetical and delicate balance among two or
more of these competing astrophysical and observational effects, we
conclude, as the simplest interpretation of our results, that GRBs
accurately trace UV metrics of cosmic star formation over $3<z<5$. As
differential effects between GRB production and star formation are
robustly expected to decrease with redshift \citep{tpj15}, our finding strongly
suggests that GRBs are providing an accurate picture of star formation
processes from $z\approx 3$ out to the highest redshifts, $z\simgt
5$. GRBs can thus be relied upon in these regimes to provide an
independent check on SFR estimates from galaxy surveys, which must
necessarily extrapolate their results to faint galaxies well beyond
their detection thresholds.

% Future work

Looking ahead, this work can be strengthened in a straightforward
manner by carrying out a complete survey of GRB host galaxies over
$3<z<5$. Such a survey would address the question of publication bias
directly, and given that the present sample already yields interesting
constraints on GRB production biases and star formation models, would
be assured of providing even greater insights.

Similar work at higher redshifts, $z > 5$, is also desirable. Given
the challenges of gathering useful data at these redshifts from
ground-based facilities, this survey work will most likely be carried
out using HST.

Meanwhile, detected GRB host galaxies in this redshift range can be
subjected to deep multiband observations, which will lead to
galaxy-by-galaxy dust extinction and stellar population
models. Targeted spectroscopic investigations will also be useful to
provide a detailed picture of the environments and properties of the
star formation occurring within this unique and -- as we have shown --
SFR-weighted selection of high-redshift star-forming galaxies.

%%%%%%%%%%%%%%%%%%%%%%%%%%%%%%%%%%%%%%%%

\acknowledgements

The authors acknowledge astrostatistical consulting support
from E. Feigelson. 
DBF expresses appreciation to MPE Directors K. Nandra and R. Bender
for administrative support of his sabbatical stay at MPE, where this
collaboration was initiated.
PS and MT acknowledge support through the Sofja Kovalevskaja Award to 
P. Schady from the Alexander von Humboldt Foundation Germany.  
RP acknowledges co-funding through NSF grant No. AST 1414246 and 
HST-GO-13831.011-A.
SK acknowledges support by DFG grant Kl 766/16-1 and
SSc support by the Th\"uringer Ministerium f\"ur Bildung, Wissenschaft 
und Kultur under FKZ 12010-514.
SSa acknowledges support from the Bundesministerium f\"ur Wirtschaft 
and Technologie through DLR (Deutsches Zentrum f\"ur Luft- und Raumfahrt e.V.)
FKZ 50 OR 1211.
KV acknowledges support by DFG grant SA 2001/2-1, and
CD acknowledges support through EXTraS, funded from the European Union's 
Seventh Framework Programme for research, technological development and 
demonstration under grant agreement no 607452.
AC is grateful for support of his visit to MPE Garching. 
Partial funding for GROND (hardware and personnel) was generously
granted from the Leibniz-Prize to Prof.\ G. Hasinger (DFG grant HA
1850/28-1).

%%%%%%%%%%%%%%%%%%%%%%%%%%%%%%%%%%%%%%%%

%\begin{thebibliography}
%\end{thebibliography}

\bibliographystyle{apj_8}
\bibliography{journals_apj,grbhosts}

%%%%%%%%%%%%%%%%%%%%%%%%%%%%%%%%%%%%%%%%%%%%%%%%%%%%%%%%%%%%%%%%%%%%%%

\begin{deluxetable}{llcclccc}

  \tablewidth{0pt}
  \tabletypesize{\footnotesize} % footnotesize,scriptsize
  \tablecolumns{7}

  \tablecaption{GRB Host Galaxy Observations}

  \tablehead{
    \colhead{GRB}  &
    \colhead{$z^a$}  &
     \colhead{Tel./Inst.}  &
    \colhead{Filt}  &
    \colhead{Exp (s)}  &
    \colhead{Position} &
    \colhead{Aper.}  &
    \colhead{mag$^b$}
%    \colhead{Notes}
  }

\startdata

 080810  & 3.355 & VLT/FORS2  &$R$ &  \nbytx{7}{1140}& 
    23\h47\m10\fss51 +00\deg19\amin11\farcs6& 1\farcs60 &  $23.40\pm0.05$ \\
 090313  & 3.375 & 2.2m/GROND &$\rp$& \nbytx{20}{369}& & & $>24.6$  \\
 090516  & 4.109 & 2.2m/GROND &$\ip$& \nbytx{20}{369}& 
    09\h13\m02\fss59 --11\deg51\amin14\farcs9& 1\farcs50  & $25.20\pm0.40$ \\
 090519  & 3.85  & VLT/FORS2  &$R$ &  \nbytx{8}{1140}&
    &   & $>27.1$  \\
%          &       & &    &                            & $>$27.9 & ~ \\     
 091109  & 3.076 & VLT/FORS2  &$R$ &  \nbytx{6}{1140}&
   20\h37\m01\fss80 --44\deg09\amin29\farcs5& 1\farcs25 & $26.06\pm0.09$ \\
 100513A & 4.772 & VLT/FORS2  &$I$ &  \nbytx{48}{240}&
    11\h18\m26\fss81 +03\deg37\amin39\farcs9& 0\farcs70 & $26.54\pm0.29$ \\
 100518A & $4.0^{+0.3}_{-0.5}$ 
                 & VLT/FORS2  &$R$ &  \nbytx{11}{1140}& & & $>28.7$ \\
 % 110818A: Corrected from R=23.96(8), I=23.49(7)
 110818A & 3.36  & VLT/FORS2  &$R$ &  \nbytx{5}{600} &
    21\h09\m20\fss94 -63\deg58\amin52\farcs4& 1\farcs50 & $24.14\pm0.05$ \\
         &       & VLT/FORS2  &$I$ &  \nbytx{12}{240}&
    21\h09\m20\fss94 --63\deg58\amin52\farcs4& 1\farcs50 & $23.40\pm0.07$ \\
 120805A & $3.1\pm0.2$  & 2.2m/GROND &$\rp$& \nbytx{12}{369}& 
    14\h26\m09\fss13 +05\deg49\amin31\farcs8& 1\farcs10  & $24.10\pm0.10$ \\
 120909A & 3.93  & VLT/FORS2  &$I$ &  \nbytx{33}{240}& 
    18\h22\m56\fss72 --59\deg26\amin54\farcs1& 0\farcs76 & $24.95\pm0.12$ \\
 120922A & $3.1\pm0.2$   
                 & 2.2m/GROND &$\rp$& \nbytx{12}{369}& &  & $>25.4$ \\
 121201A & 3.385 & 2.2m/GROND &$\rp$& \nbytx{18}{369}&
    00\h53\m52\fss16 --42\deg56\amin35\farcs1& 1\farcs50 & $24.95\pm0.21$ \\
 130408A & 3.757 & 2.2m/GROND &$\rp$& \nbytx{18}{369}& &  & $>25.9$  \\
         &       & 2.2m/GROND &$\ip$& \nbytx{18}{369}& &  & $>25.2$  \\
  \enddata
  \tablecomments{$^a$ Redshifts with errors are photometric redshifts.
                 $^b$ Upper limits are at the 2$\sigma$ confidence level.}
  \label{tab:obs}
\end{deluxetable}

%%%%%%%%%%%%%%%%%%%%%%%%%%%%%%%%%%%%%%%%%%%%%%%%%%%%%%%%%%%%%%%%%%%%%%

\begin{deluxetable}{lcccccccl}

  % Table width is calculated when you set to ``0pt''. Then you reset
  % it here to the maximum across all table pages: 
  \tablewidth{611pt}
  \rotate
  \tabletypesize{\scriptsize}
  \tablecolumns{9}

  \tablecaption{GRB Host Galaxies over $3.0 < z < 5.0$}

  \tablehead{
    \colhead{GRB}  &
    \colhead{$z$}  &
    \colhead{mag}  &
    \colhead{$E_{\rm B-V}$}  &
    \colhead{DM} & 
    \colhead{$\beta$} & 
    \colhead{\Mrfuv} &
    \colhead{SFR (\Msunpyr)} &
    \colhead{References}
  }

\startdata

  %%%%%%%%%%%%%%%%%%%%%%%%%%%%%%%%%%%%%%%%%%%%%%%%%%%
  % R/r-band or I/i-band at 3 < z < 3.5

  140114A & $3.0^{+0.3}_{-0.3}$ & $R_{\rm AB}=24.4(2)$ & 0.014 & $47.03^{+0.24}_{-0.28}$ 
          & $-1.67(2)$ & $\bmm{-21.15(20)(22)}$ &  $24.5(49)^{+6.0}_{-5.3}$
          & \citet{kmf+15} \\
  080607  & 3.036 & $\rp=26.75(46)$ & 0.019 & 47.06 & $-$1.92 & $\bmm{-18.84(46)}$ &  $1.9(10)$
          & \citet{plt+13} \\
          &       & $f_{\rm 3.6} = 2.53(24)$\,\uJy & &  &$-$1.59 & $-$21.88(10) & 55.5(53)
          & \citet{chary07} \\
  060607A & 3.075 & $R>28.36$     & 0.025 & 47.09 &$-$2.10& $>-$17.26  & $<$0.3
          & \citet{hmj+12} \\
          &       & $m_{\rm 775w}>30.15$ &   &      & $-$2.29& $\bmm{>-15.52}$ & $<$0.1 
          & \cite{svth11} \\
          &       & $H_{\rm AB}>26.5$ &     &       &$-$2.00& $>-$19.08  & $<$2.0
          & \citet{cpp+09} \\
  091109  & 3.076 & $\rp=26.06(9)$  & 0.026 & 47.09 &$-$1.84& $\bmm{-19.58(9)}$ & 4.2(4)
          & This work \\
  120922A & $3.1^{+0.2}_{-0.2}$ & $\rp>25.4$     & 0.128 & $47.11^{+0.16}_{-0.17}$ 
          & $-$2.00& $\bmm{>-20.50(12)}$ & $<$7.5(8)
          & This work \\
  120805A & $3.1^{+0.3}_{-0.3}$ & $\rp=24.1(1)$ & 0.027 & $47.11^{+0.24}_{-0.26}$ 
          & $-1.62(3)$ & $\bmm{-21.57(10)(22)}$ & $39.0(38)^{+9.5}_{-8.3}$
          & This work \\
  111123A & 3.152  & $I=23.55$      & 0.047 & 47.16 &$-$1.57& $\bmm{-22.05(3)}$  & 67(21)
          & \citet{gcn.14273} \\
  020124 & 3.198 & $R_{\rm AB}>30.0$ & 0.041 & 47.19 &$-$2.27& $\bmm{>-15.72}$ & $<$0.06
          & \citet{bkb+02, cpp+09} \\
          &       & $H_{\rm AB}>26.1$ &      &       &$-$2.00& $>-$19.56 & $<$3.1
          & \citet{cpp+09} \\
  060926  & 3.206 & $f_{3.6}=1.65(7)$\,\uJy & 0.138 & 47.20 &$-$1.62& $\bmm{-21.59(5)}$ & 40.1(17)
          & \citet{lbc11} \\
  060526  & 3.221 & $R>27.46$      &  0.056 & 47.21 &$-$2.00& $>-$18.32    & $<$1.0
          & \citet{hmj+12} \\
          &       & $m_{\rm 775w}>28.41$ &   &      & $-$2.09& $\bmm{>-17.36}$ & $<$0.4 
          & \cite{svth11} \\
  050319  & 3.24  & $f_{3.6}=0.80(9)$\,\uJy & 0.010 &47.23 &$-$1.69& $\bmm{-20.95(12)}$ & 19.7(22)
          & \citet{lbc11} \\
  050908  & 3.347 & $R=27.76(45)$  & 0.021 & 47.31 &$-$2.01& $\bmm{-18.00(45)}$ & 0.7(4)
          & \citet{hmj+12} \\
          &       & $H_{\rm AB}>26.0$ &    &      &$-$2.00& $>-$19.73    & $<$3.7
          & \citet{cpp+09} \\
  080810  & 3.355 & $\rp=23.40(5)$  & 0.024 & 47.32 &$-$1.53& $\bmm{-22.43(5)}$ & 102.7(48)
          & This work \\
  110818A & 3.36  & $R=24.14(5)$ & 0.031 & 47.32 &$-$1.61& $\bmm{-21.68(5)}$ & 44.3(21)
          & This work \\
          &       & $I=23.40(7)$ &       &      &$-$1.54& $-$22.31(7) & 89.1(59)
          & This work \\
  030323  & 3.372 & $m_{\rm 606w}=27.4(1)$ & 0.049 & 47.33 &$-$1.96& $\bmm{-18.47(10)}$ & 1.2(1)
          & \citet{cpp+09} \\
  090313  & 3.375 & $\rp>24.6$  & 0.024 & 47.33 &$-$2.00& $\bmm{>-21.19}$ & $<$14.2
          & This work \\
  121201A & 3.385 & $\rp=24.95(21)$ & 0.008 & 47.34 &$-$1.70& $\bmm{-20.84(21)}$ & 17.4(37)
          & This work  \\
  971214  & 3.418 & $R=25.60(15)$ & 0.016 & 47.37 &$-$1.77& $\bmm{-20.21(15)}$ & 8.6(13) 
          & \citet{kdr+98, plt+13} \\
  060707  & 3.424 & $R=24.86(6)$    & 0.019 & 47.37 &$-$1.69& $\bmm{-20.97(6)}$ & 19.9(11)
          & \citet{hmj+12} \\
          &       & $f_{\rm 3.6} = 1.10(10)$\,\uJy & & &$-$1.65& $-$21.34(9) & 30.2(27)
          & \citet{lbc11} \\
  061110B & 3.434 & $R=26.04(29)$   & 0.035 &47.38 &$-$1.82& $\bmm{-19.82(29)}$ & 5.5(17)
          & \citet{hmj+12} \\ 
  980329  & $3.5^{+0.4}_{-0.4}$ & $m_{\rm 7228lp}=26.2(1)$ & 0.064 & $47.43^{+0.28}_{-0.32}$ 
          &$-$1.82(3)& $\bmm{-19.73(10)(23)}$ & 5.1(5)(13)
          & \citet{jah+03} \\ 
          &        & $I=26.28(27)$  &    &  &$-$1.84(2)& $-$19.59(27)(22) & 4.3(12)(11)
          & \citet{bkd+02} \\ \tableline

  % R/r-band or I/i-band at z < 3.5
  %%%%%%%%%%%%%%%%%%%%%%%%%%%%%%%%%%%%%%%%%%%%%%%%%%%
  % I/i-band preferred at 3.5 < z < 4.8

  060115  & 3.533 & $R=27.14(53)$   & 0.113 & 47.45 &$-$1.91& $\bmm{-18.94(53)}$ & 2.1(13)
          & \citet{hmj+12} \\

  090323  & 3.569 & $\rp=24.87(15)$ & 0.021 & 47.48 &$-$1.67& $-$21.17(15)$^*$ & 25.0(37)
          & \citet{mkr+10} \\
          &       & $\ip=24.25(18)$ &    &  &$-$1.62& $\bmm{-21.60(18)}$ & 40.4(73)
          & \citet{mkr+10} \\

  070721B & 3.626 & $R=27.53(44)$   & 0.027 & 47.52 &$-$1.97& $\bmm{-18.39(44)}$ & 1.1(6)
          & \citet{hmj+12} \\
          &       & $H_{\rm AB}>25.8$    &   &  &$-$2.00& $>-$20.07  & $<$5.0
          & \citet{cpp+09} \\

  060906 & 3.686  & $f_{\rm 3.6} < 0.28$\,\uJy & 0.175 & 47.56 &$-$2.00& $\bmm{>-20.60}$ & $<$8.2
          & \citet{lbc11} \\

  130408A & 3.757 & $\rp>25.9$& 0.220 & 47.61 &$-$2.00& $>-$20.77$^*$ & $<$9.6
          & This work \\
          &        & $\ip>25.2$&    &    &$-$2.00& $\bmm{>-21.13}$ & $<$13.4
          & This work \\

  060605  & 3.773 & $R=26.4(3)$ & 0.044 & 47.62 &$-$1.82& $-$19.81(30)$^*$ & 5.5(17)
          & \citet{fkk+09} \\
          &       & $m_{\rm 775w}=28.07(20)$ &   &      & $-$2.02& $\bmm{-17.94(20)}$ & 0.7(1)
          & \cite{svth11} \\

  050502  & 3.793 & $f_{\rm 3.6} < 0.21$\,\uJy & 0.009 & 47.64 &$-$2.00& $\bmm{>-20.34}$ & $<$6.5
          & \citet{lbc11} \\

  081029  & 3.848 & $R_{\rm AB}>26.3$ & 0.027 & 47.67 &$-$2.00& $\bmm{>-19.87^*}$ & $<$4.2
          & \citet{ngk+11} \\

  090519  & 3.85  & $R>27.1$   & 0.035  & 47.67 &$-$2.00& $\bmm{>-19.09^*}$ & $<$2.0
          & This work \\

  060210  & 3.913 & $I_{\rm AB}=24.40(20)$ & 0.080 & 47.72 &$-$1.61 & $\bmm{-21.70(20)}$ & 45.5(92)
          & \citet{pcb+09} \\
          &       & $f_{\rm 3.6} = 1.41(10)$\,\uJy & & &$-$1.60& $-$21.80(7) & 50.4(36)
          & \citet{lbc11} \\

  120909A & 3.93  & $I = 24.95(12)$  &  0.076 & 47.73 &$-$1.67 & $\bmm{-21.16(12)}$ & 24.7(29)
          & This work \\

  050730  & 3.968 & $\ip > 26.6$  & 0.043 & 47.75 &$-$2.00& $>-$19.49 & $<$3.0
          & \citet{cpp+09} \\
          &       & $m_{\rm 775w}>28.88$ &   &      & $-$2.10& $\bmm{>-17.21}$ & $<$0.3 
          & \cite{svth11} \\

  100518A & $4.0^{+0.3}_{-0.5}$ & $R>28.7$ & 0.067 & $47.77^{+0.19}_{-0.34}$ 
          & $-2.05(5)$ & $\bmm{> -17.71^*}\bmm{^{+0.47}_{-0.35}}$ & $< 0.5(3)$
          & This work \\ 

  060206  & 4.048 & $m_{\rm 814W}=27.6(1)$ & 0.011 & 47.80 &$-$1.96& $\bmm{-18.47(10)}$ & 1.2(1)
          & \citet{cpp+09} \\

  090516  & 4.109 & $\ip=25.2(4)$   & 0.050  & 47.84 &$-$1.69& $\bmm{-20.99(40)}$ & 20.4(91)
          & This work \\

  080916C & $4.35^{+0.15}_{-0.15}$ & $I>26$ & 0.291 & $47.99(9)$ 
          & $-$2.00 & $\bmm{>-20.60(6)}$ & $<$8.2(5)
          & \citet{gck+09} \\

  060223A & 4.406 & $R>26.69$ & 0.100 & 48.02 &$-$2.00& $>-$20.25$^*$ & $<$5.9
          & \citet{hmj+12} \\
          &       & $m_{\rm 110w}=27.96(15)$ &  &  & $-$1.98& $\bmm{-18.33(15)}$ & 1.1(2) 
          & \cite{svth11} \\

  000131 & 4.50  &  $R>25.7$  & 0.047 & 48.08 &$-$2.00& $>-$21.19$^*$ & $<$14.2
          & \citet{ahp+00} \\
          &       & $I>24.85$ &   &  &$-$2.00& $\bmm{>-21.44}$ & $<$17.9
          & \citet{ahp+00} \\

  090205  & 4.65  & $I_{\rm AB}=25.22(13)$ & 0.101  & 48.16 &$-$1.66 & $\bmm{-21.26(13)}$ & 27.7(35)
          & \citet{dpf+10} \\
          &        & $J_{\rm AB}>25.8$ &   &   &$-$2.00& $>-$20.57 & $<$8.0
          & \citet{dpf+10} \\

  100219A & 4.667 & $\ip=26.7(5)$ & 0.066 & 48.17 &$-$1.82& $\bmm{-19.74(50)}$ & 5.1(30)
          & \citet{tfg+13} \\

  100513A & 4.772 & $I=26.54(29)$  & 0.046 & 48.23 &$-$1.81 & $\bmm{-19.88(29)}$ & 6.0(18)
          & This work \\ \tableline

  % I/i-band preferred at 3.5 < z < 4.8
  %%%%%%%%%%%%%%%%%%%%%%%%%%%%%%%%%%%%%%%%%%%%%%%%%%%
  % z-band preferred at z > 4.8

  060510B & 4.942 &  $f_{\rm 3.6} = 0.23(4)$\,\uJy & 0.034 &48.31 &$-$1.74 & $\bmm{-20.51(17)}$ & 12.0(21)
          & \citet{chary07} \\

  111008A & 4.99  & $\zp>25.6$ & 0.004 & 48.34 &$-$2.00& $\bmm{>-20.80}$ & $<$9.9
          & \citet{shk+14} \\

  %%%%%%%%%%%%%%%%%%%%%%%%%%%%%%%%%%%%%%%%%%%%%%%%%%%

  \enddata

  \tablenotetext{*}{Observations will be affected by
    absorption at and blueward of the Lyman-alpha transition at the
    GRB redshift; see table notes for details.}

  \tablecomments{
   (1) The third column contains measured magnitudes/fluxes, i.e. without any
       correction for Galactic foreground or host-intrinsic extinction.
       Observed magnitudes $RIJHK$ are in the Vega system, and $\rp\ip\zp$
       magnitudes and HST measurements are in the AB system, except
       where noted.  
   (2) Spitzer upper limits (3$\sigma$) from \citet{lbc11} are used
       only when no competitive limits are available at other wavelengths.
   (3) All upper limits are 2$\sigma$ confidence level; 
       in some cases, reported 3$\sigma$ upper limits have been transformed
       into 2$\sigma$ limits by adding 0.44 mag (e.g. for measurements from
       \citealt{hmj+12}).
       Observed magnitudes are corrected for Galactic foreground
       extinction \citep{schlafly+11} before conversion to absolute magnitudes,
       using $A_V = 3.1 E(B-V)$ and 
       $A_r (A_i) = 0.80 (0.61) \times A_V$, $A_R (A_I) = 0.75 (0.48)
       \times A_V$, $A_J (A_H) = 0.29 (0.18) \times A_V$, $A_{\rm 606w} =
       0.90 A_V$, $A_{\rm 723lp} = 0.71 A_V$, $A_{\rm 775w} = 0.63 A_V$, $A_{\rm 814W} = 0.58
       A_V$, $A_{\rm 110w} = 0.35 A_V$, and $A_{3.6} = 0.046 A_V$. 
   (4) $\beta$ is the slope of the assumed power-law spectrum
       of the star-forming galaxy ($f_\lambda \propto \lambda^\beta$)
       which is  used to compute the k-correction according to
       $k = 2.5 (1+\beta) \log(1+z)$.
   (5) \Mrfuv\ is the inferred absolute AB magnitude of the host
       galaxy (or limit) at $\lambda=1600$\AA\ in the rest
       frame. Measurements and limits used to construct the
       \Mrfuv\ luminosity function are listed in boldface. 
   (6) Horizontal lines indicate that, as discussed in the text,
       at redshifts $z > 3.5$ ($z > 4.8$), $I$/\ip\ ($z$/\zp)
       measurements are preferred, as Ly-$\alpha$ has moved into the
       $R$/\rp\ ($I$/\ip) bandpass. \Mrfuv\ estimates with an
       asterisk have been corrected for this effect assuming a 60\%
       suppression of the continuum blueward of Lyman-alpha. Applied
       corrections are: $-0.10$\,mag for 090323; $-0.21$\,mag 
       for 130408A; $-0.15$\,mag for 060605, 081029, and 090519;
       0.00, $-0.24$, and $-0.46$\,mag for 100518A at $z=3.5$, 4.0,
       and 4.3, respectively; $-0.52$\,mag for 060223A; and
       $-0.56$\,mag for 000131. Quoted $\beta$ values and SFR
       estimates derived from these measurements incorporate these
       corrections. }  

  \label{tab:hosts}

\end{deluxetable}

%%%%%%%%%%%%%%%%%%%%%%%%%%%%%%%%%%%%%%%%%%%%%%%%%%%%%%%%%%%%%%%%%%%%%%

\end{document}